\documentclass[a4paper,12pt]{article}

\pdfoutput=1

\usepackage[utf8]{inputenc}
\usepackage{amsmath}
\usepackage{amssymb}
\usepackage{amsfonts}
\usepackage{mathrsfs}
\usepackage{graphicx}
\usepackage{booktabs,adjustbox}
\usepackage{caption}
\usepackage{slashed}
\usepackage{verbatim}
\usepackage{float}
\usepackage{setspace}
\usepackage{subfig}
\usepackage{jheppub}

\usepackage{color}
\usepackage{ulem}

\usepackage{bookmark}

\makeatletter
\newcommand{\thickhline}{%
	\noalign {\ifnum 0=`}\fi \hrule height 1pt
	\futurelet \reserved@a \@xhline
}
\makeatother

\allowdisplaybreaks

\title{\boldmath Freeze-in Dirac neutrinogenesis: thermal leptonic CP asymmetry}

\author[a]{Shao-Ping Li,}
\author[a,1]{Xin-Qiang Li,\note{Corresponding author.}}
\author[a]{Xin-Shuai Yan}
\author[a]{and Ya-Dong Yang}

\affiliation[a]{Institute of Particle Physics and Key Laboratory of Quark and Lepton Physics~(MOE),\\Central China Normal University,Wuhan, Hubei 430079, China}

\emailAdd{ShowpingLee@mails.ccnu.edu.cn}
\emailAdd{xqli@mail.ccnu.edu.cn}
\emailAdd{xinshuai@mail.ccnu.edu.cn}
\emailAdd{yangyd@mail.ccnu.edu.cn}

\abstract{We present a freeze-in realization of the Dirac neutrinogenesis in which the decaying particle that generates the lepton-number asymmetry is in thermal equilibrium. As the right-handed Dirac neutrinos are produced non-thermally, the lepton-number asymmetry is accumulated and partially converted to the baryon-number asymmetry via the rapid sphaleron transitions. The necessary CP-violating condition can be fulfilled by a purely thermal kinetic phase from the wavefunction correction in the lepton-doublet sector, which has been neglected in most leptogenesis-based setup. Furthermore, this condition necessitates a preferred flavor basis in which both the charged-lepton and neutrino Yukawa matrices are non-diagonal. To protect such a proper Yukawa structure from the basis transformations in flavor space prior to the electroweak gauge symmetry breaking, we can resort to a plethora of model buildings aimed at deciphering the non-trivial Yukawa structures. Interestingly, based on the well-known tri-bimaximal mixing with a minimal correction from the charged-lepton or neutrino sector, we find that a simultaneous explanation of the baryon-number asymmetry in the Universe and the low-energy neutrino oscillation observables can be attributed to the mixing angle and the CP-violating phase introduced in the minimal correction.}

\begin{document}
	\maketitle
	\flushbottom
	
\section{Introduction}
\label{sec:intro}
		
Recent developments in particle physics and cosmology, especially those related to the neutrino mass, dark matter, as well as baryon asymmetry of the Universe (BAU), have highlighted the importance of feeble couplings. Actually, feeble couplings are already present in the Yukawa couplings of the light charged fermions within the Standard Model (SM); \textit{e.g.}, the SM predicts an electron Yukawa coupling with $y_e\simeq 10^{-6}$. If one also accepts feeble Yukawa couplings of the Dirac neutrinos, the smallness of neutrino masses can then be simply addressed via the Higgs-like mechanism with three right-handed Dirac neutrino singlets. Feeble couplings can also play an important role in the early Universe. As a specific example, the feebleness allows for a freeze-in production of the dark matter abundance, which can be effectively kept from large annihilation~\cite{Hall:2009bx,Bernal:2017kxu}. Moreover, the feebleness stirs up a new leptogenesis, named Dirac neutrinogenesis (DN)~\cite{Dick:1999je}, in which the out-of-equilibrium condition for generating the lepton-number~($L$) asymmetry can be guaranteed and the baryon-number~($B$) asymmetry is generated via thermal sphaleron transitions~\cite{Kuzmin:1985mm}, even in a theory with $B-L=0$ initially. 

In the typical versions of DN mechanism~\cite{Dick:1999je,Murayama:2002je,Cerdeno:2006ha,Gu:2007mc,Bechinger:2009qk,Narendra:2017uxl}, the lepton-number asymmetry is generated by heavy particle decays with a non-thermal distribution. In addition, in order to discuss the loop correction for nonzero kinetic phase (or the absorptive part of the decay amplitude), one usually focuses on the new particle sector, which is also the general case in seesaw-based leptogenesis~\cite{Giudice:2003jh,Buchmuller:2004nz,Davidson:2008bu}, while the contribution from wavefunction correction in the lepton-doublet sector has not yet been considered to the best of our knowledge. 

There could be two possible reasons for having neglected the lepton-doublet wavefunction contribution. On the one hand, the charged-lepton flavors are widely assumed to populate in the diagonal basis, and thus the leptonic CP asymmetry cannot be generated from the self-energy diagrams in the lepton-doublet sector. Interestingly, however, it has been pointed out earlier that the well-known tri-bimaximal (TB) mixing pattern~\cite{Harrison:2002er}, with a minimal correction from the charged-lepton or neutrino sector, can produce compatible neutrino oscillation data while retaining its compelling prediction~\cite{Albright:2008rp,He:2011gb}. In this respect, a nontrivial combination of the charged-lepton and neutrino mixings is preferred to produce the oscillation observables. On the other hand, even with a non-diagonal charged-lepton Yukawa matrix, there is no on-shell cut in the self-energy loop at zero-temperature regime, and hence no CP asymmetry either. Nevertheless, it has been illustrated in ref.~\cite{Giudice:2003jh} and later implemented in ref.~\cite{Hambye:2016sby} that, at high-temperature regime where thermal effects come into play, the zero-temperature cutting rules should be superseded by the thermal cuts~\cite{Das1997}, allowing consequently nonzero contributions to the leptonic CP asymmetry that would otherwise vanish at the vacuum regime.

Therefore, as will be exploited in this paper, when both thermal effects and nontrivial mixings in the charged-lepton and neutrino sectors are taken into account, one can expect the leptonic CP asymmetry at finite temperature to carry a nonzero imaginary piece, \textit{i.e.}, $\text{Im}[(Y_\nu Y^\dagger_\nu)(Y_\ell Y^\dagger_\ell)]\neq0$, where $Y_{\ell}$ and $Y_{\nu}$ denote respectively the charged-lepton and neutrino Yukawa matrices that are responsible for their respective masses and mixings. This enables us to exploit a direct interplay between the BAU and the neutrino oscillation observables in a minimal setup, without tuning additional Yukawa couplings beyond $Y_{\ell,\nu}$. Furthermore, since the feeble neutrino Yukawa couplings essentially prompt an out-of-equilibrium condition (\textit{i.e.}, the right-handed Dirac neutrinos undergo a freeze-in production in the early Universe), there is no need to invoke much heavier dynamical degrees of freedom (d.o.f), and the evolution of the lepton-number asymmetry can be much simplified as well.

The remainder of this paper is organized as follows. We begin in section~\ref{sec:2} with a brief overview of the DN mechanism, and then calculate the leptonic CP asymmetry with two different thermal cuts in a model-independent way. The Boltzmann equation for the evolution of the lepton-number asymmetry in the freeze-in regime is also derived here. In section~\ref{sec:3}, we discuss, from a phenomenological perspective, some minimal corrections to the well-known TB mixing pattern, which are subsequently found to be related with the generated baryon-number asymmetry. In section~\ref{sec:4}, we identify the scalars participating in the out-of-equilibrium decay and perform our detailed numerical analyses. Our conclusions are finally made in section~\ref{sec:con}. 

\section{Thermal leptonic CP asymmetry and evolution}
\label{sec:2}

\subsection{Dirac neutrinogenesis}

The basic idea of DN~\cite{Dick:1999je} can be summarized as follows. In a theory without lepton-number-violating Lagrangian, due to the feeble neutrino Yukawa couplings that prevent the left- and right-handed Dirac neutrinos from equilibration (the case when they are in equilibration will be dubbed as the ``L-R equilibration" from now on), the leptonic CP asymmetry from a heavy particle decay in the early Universe can result in a net lepton-number asymmetry stored in the right-handed Dirac neutrinos and the lepton doublets. As the sphaleron transitions act only on the left-handed particles, the net lepton-number asymmetry stored in the lepton doublets will be partially converted to the baryon-number asymmetry via rapid sphaleron processes, while the portion stored in the right-handed Dirac neutrinos keeps intact. After the sphaleron freezes out around $T\simeq \mathcal{O}(100)$~GeV, a net baryon-number (as well as lepton-number) asymmetry survives till today. 

Since all the SM species (except for the added right-handed neutrinos) keep in chemical equilibrium during the thermal sphaleron epoch, $10^2~\text{GeV}<T<10^{12}~\text{GeV}$, the final baryon-number asymmetry can be determined at the sphaleron decoupling temperature via the following conversion relations: 
\begin{align}\label{sphaleron relation}
Y_{\Delta B}=c\,Y_{\Delta(B- L_{SM})}=c\,Y_{\Delta L_{\nu_R}},
\end{align}
where $L_{SM}=L-L_{\nu_R}$ represents the lepton number in the SM sector, \textit{i.e.}, the total lepton number ($L$) minus the portion in the right-handed neutrino part ($L_{\nu_R}$). Note that the first relation results from the thermal sphaleron transition~\cite{Harvey:1990qw}, with the coefficient given by $c=(8N_f+4N_H)/(22N_f+13N_H)$, where $N_f$ and $N_H$ denote the numbers of fermion generations and Higgs doublets, respectively; while the second one is essentially due to the conservation law, $B-L\equiv B-L_{SM}-L_{\nu_R}$, which applies at the sphaleron decoupling temperature, and will be used to determine the final baryon-number asymmetry $Y_{\Delta B}$. 

In order to prompt the necessary out-of-equilibrium condition so that the generated lepton-number asymmetry is kept from being washed out by the L-R equilibration, the lepton-number-violating thermal decay rate must be sufficiently smaller than the expansion rate of the Universe, which typically requires the Dirac neutrino Yukawa couplings to be $Y_\nu\lesssim \mathcal{O}(10^{-8})$. Such feeble couplings, in spite of their non-aesthetic nature, are generically present, if the sub-eV Dirac neutrino masses are generated by the Higgs-like mechanism with a vacuum expectation value (VEV) chosen around the electroweak scale. On the other hand, such a mechanism of neutrino mass generation is often criticized on account of \textit{naturalness}, and dynamical explanations of the smallness of neutrino masses are, therefore, more biased by enlarging the Yukawa space and/or introducing sufficiently heavy particles, which have also been considered in explicit realizations of the  DN~\cite{Murayama:2002je,Cerdeno:2006ha,Gu:2007mc,Bechinger:2009qk,Narendra:2017uxl}.

Nonetheless, for these dynamical explanations with overabundant Yukawa parameters, reliable phenomenological predictions rely on the particular bases and values of the unknown Yukawa couplings beyond those that can be directly fixed by the lepton flavor spectrum. In particular, a simple connection between the BAU and the low-energy neutrino oscillation observables cannot be established, if the DN realization has nothing to do with the Yukawa couplings that are directly responsible for the lepton masses and mixings. Furthermore, it is difficult, if not impossible, to detect the additional, sufficiently heavy particles at current colliders.     

In this paper, as an underlying theory for explaining the feebleness of the Yukawa couplings for both light charged leptons and neutrinos is still unknown, if it were to exist, we shall take the feeble neutrino couplings as a starting point. In this context, we provide a new DN realization in which the decaying particle that generates the lepton-number asymmetry is in thermal equilibrium. For this purpose, we consider a Higgs-like doublet which has a vacuum mass around $\mathcal{O}(10^2)$~GeV and feeble couplings to the right-handed Dirac neutrinos. The feeble Yukawa couplings ensure that the right-handed Dirac neutrinos never reach equilibrium with the thermal bath. In addition, the leptonic CP asymmetry is induced by the self-energy correction in the lepton-doublet sector due to the thermal effects. This realization allows us to establish a simple connection between the BAU and the low-energy neutrino oscillation observables, and, at the same time, renders the detection of the scalars \textit{at least} in principle possible at current and future colliders.  

\subsection{Theoretical setup}

In this subsection, we shall adopt a real-time formalism in thermal field theory to calculate the thermal leptonic CP asymmetry. To appreciate the subtlety in calculating the CP asymmetry between thermal field theory and non-equilibrium quantum field theory (QFT), we shall use two different thermal cuts and compare the corresponding consequences arising from the different dependence on the distribution functions. The evolution of the lepton-number asymmetry will be determined by a simplified Boltzmann equation in the freeze-in regime. 

\subsubsection{Thermal field theory: real-time formalism}

There are two equivalent approaches in thermal field theory, the real-time and the imaginary-time formalism~\cite{Landsman:1986uw,Das1997}. Within the real-time formalism, we do not need to perform analytic continuation for the physical region, but there is a doubling of d.o.f dual to each field presented in vacuum QFT. As a result, the interaction vertices are doubled, and the thermal propagators have a $2\times 2$ structure. In the following, we shall adopt this formalism to calculate the thermal leptonic CP asymmetry. 

\begin{figure}[t]
	\centering	
	\includegraphics[width=0.80\textwidth]{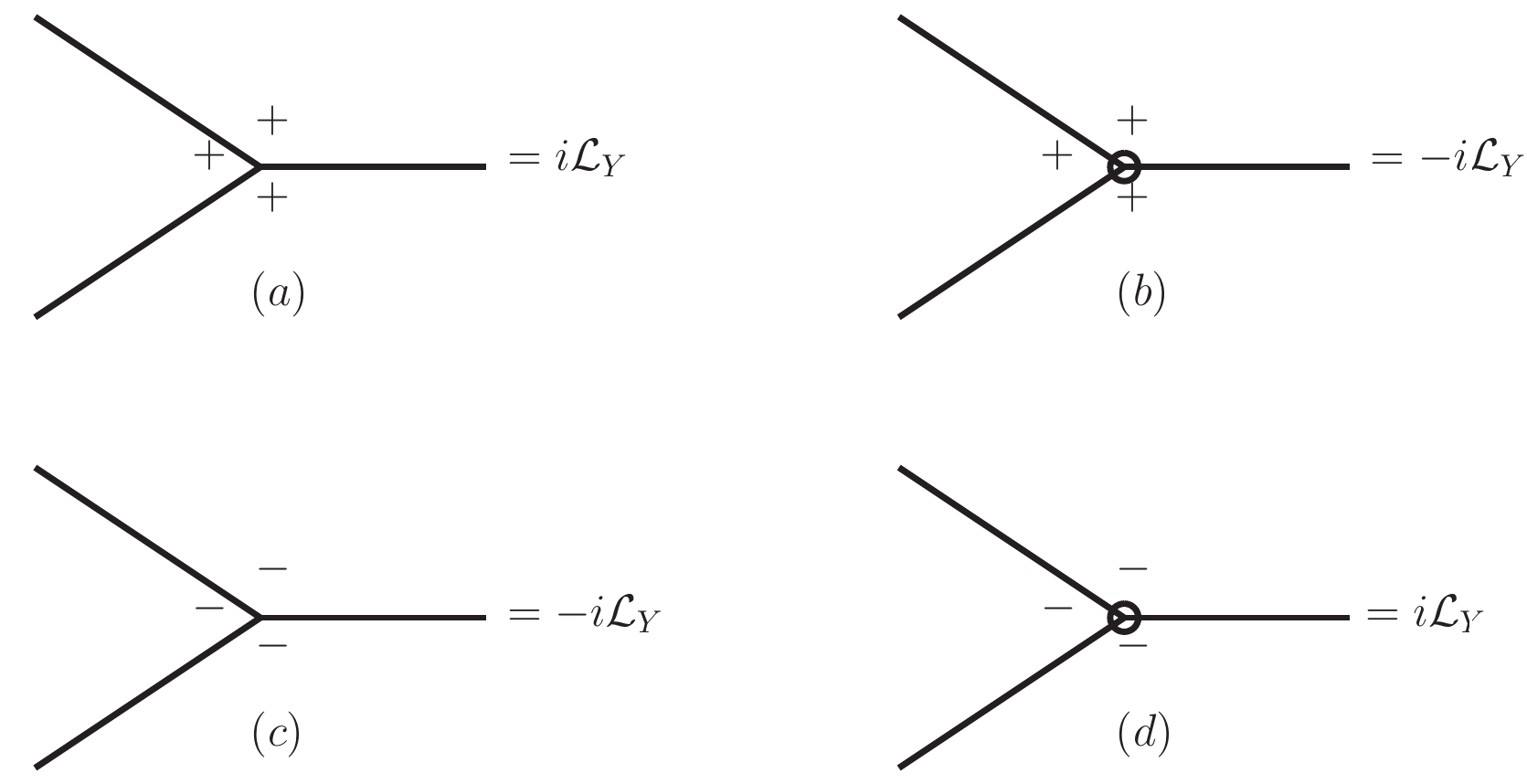}
	\caption{\label{thermal FeynRule} Circling rules in doubled interaction vertices specified by different thermal indices $\pm$. Here $\mathcal{L}_{Y}$ can be either the Yukawa Lagrangian of the SM extended by a neutrino term or of the neutrinophilic two-Higgs-doublet model (2HDM), to be discussed later.}
\end{figure}

\begin{figure}[t]
	\centering	
	\includegraphics[scale=0.80]{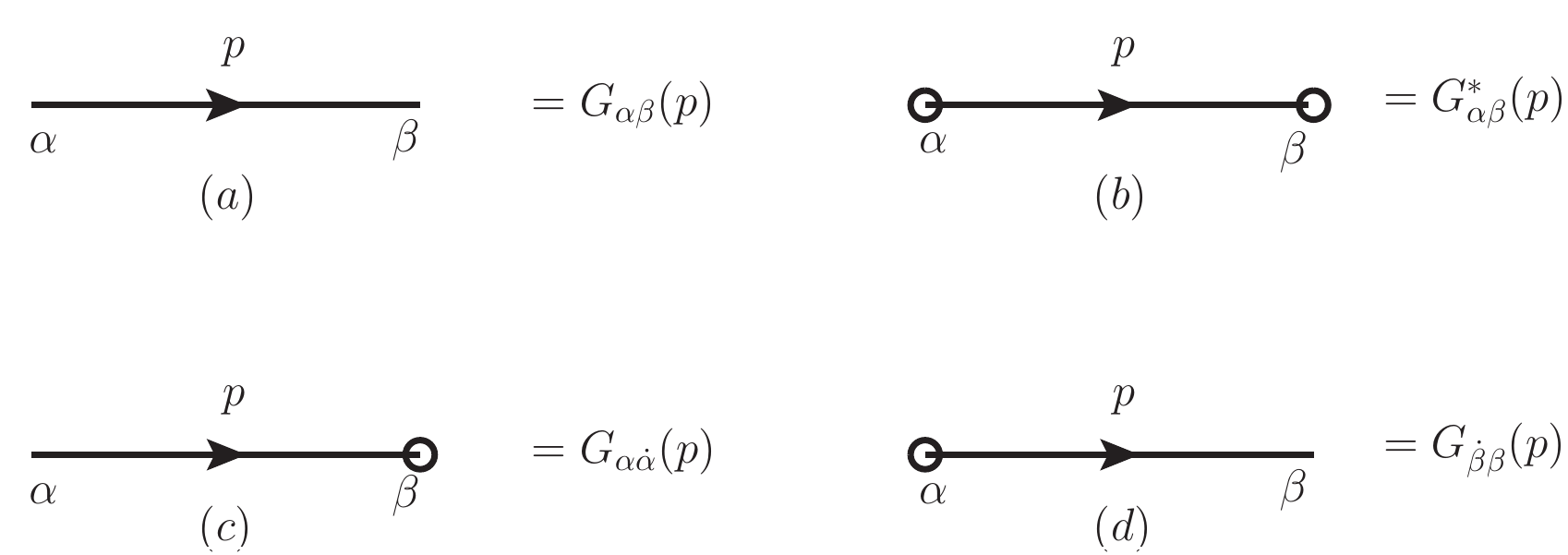}
	\caption{\label{thermal FeynRule2} Circling rules in thermal propagators. Here $\alpha$ and $\beta$ take the thermal indices $\pm$. Note that the propagator indices in $(c)$ and $(d)$ are completely determined by the uncircled ones, with $\dot{\alpha}$ and $\dot{\beta}$ taking the opposite signs of $\alpha$ and $\beta$, respectively.}
\end{figure}

Within the real-time formalism, while both the closed-time path formulation and the thermo-field dynamics can be used, we shall follow here the former~\cite{Das1997}. In this formulation, the circling rules necessary for evaluating the absorptive part of the decay amplitude are given in figures~\ref{thermal FeynRule} (for the interaction vertices) and \ref{thermal FeynRule2} (for the thermal propagators). To get a compact expression for the thermal propagators and a unified rule in writing the amplitude for each vertex, we adopt a convention in which the numerator factor $\slashed p \pm m$ of the fermion propagator is decomposed into a spin summation $\sum_s u^s \bar u^s (v^s \bar v^s)$, where the Dirac spinors would then be attached to each vertex. Thus, the thermal propagators, with the subscript indices $\pm$ specifying the corresponding matrix elements, can be written, explicitly, as
\begin{align}
G_{++}(p)&=\frac{i}{p^2-m^2+i \epsilon}\pm 2\pi f_{B/F}(\vert p^0 \vert)\delta(p^2-m^2),\\[0.2cm]
G_{--}(p)&=\left(G_{++}(p)\right)^*,\\[0.2cm]
G_{+-}(p)&=2\pi \left[\pm f_{B/F}(\vert p^0\vert)+\theta(-p^0)\right]\delta(p^2-m^2),\\[0.2cm]
G_{-+}(p)&=2\pi \left[\pm f_{B/F}(\vert p^0\vert)+\theta(p^0)\right]\delta(p^2-m^2),
\end{align}
where $f_{B/F}(E)=(e^{E/T}\mp 1)^{-1}$ are the standard distribution functions, with $B$ and $F$ referring to the bosons and fermions, respectively. $\theta(p^0)$ denotes the Heaviside step function. Note that the circled indices in figures~\ref{thermal FeynRule2}$(c)$ and \ref{thermal FeynRule2}$(d)$ are determined completely by the uncircled ones, with $\dot{\alpha}$ and $\dot{\beta}$ taking the opposite signs of $\alpha$ and $\beta$, respectively. For example, the propagator in figure~\ref{thermal FeynRule2}$(c)$ with an uncircled thermal index $\alpha=+$ is given by $G_{+-}(p)$.

\subsubsection{Leptonic CP asymmetry: model-independent approach}

As a generic model-independent discussion, let us consider the neutrino Yukawa term,
\begin{align}\label{Yukawa}
-\mathcal{L}_\nu=Y_\nu \bar{L} \tilde{\Phi}\nu_R+ {\rm h.c.},
\end{align} 
added to the SM Lagrangian. Here we denote the lepton doublet by $L$, and assume that the Higgs doublet $\tilde{\Phi}\equiv i \sigma_2 \Phi^\ast$, with $\sigma_2$ being the Pauli matrix, does not populate well above the electroweak scale. It could be the SM Higgs doublet or a second Higgs doublet which may or may not couple to quarks. To forbid the appearance of Majorana neutrino mass term and, at the same time, to realize the DN, the right-handed neutrinos must carry a nonzero lepton number under some global $U(1)$ symmetry. After the Higgs doublet develops a non-vanishing VEV, $\langle\Phi\rangle=\left(0,v_\Phi/\sqrt{2}\right)^{T}$, the vacuum neutrino mass is then given by $m_\nu=v_\Phi Y_\nu/\sqrt{2}$.

\begin{figure}[t]
	\centering	
	\includegraphics[width=0.85\textwidth]{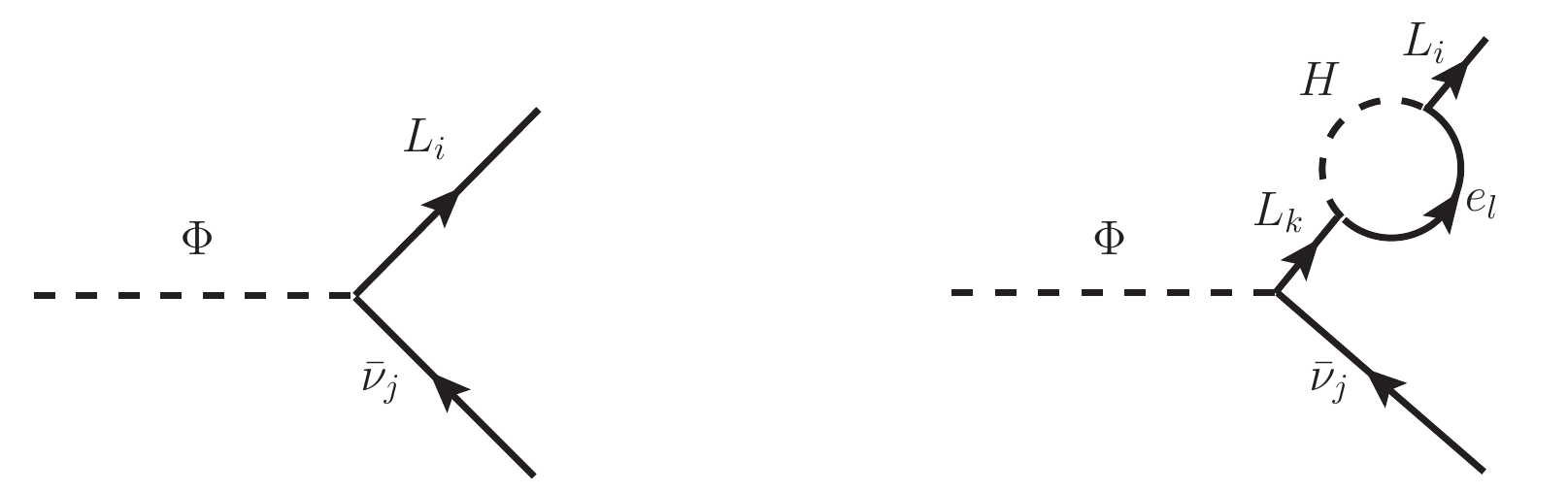}
	\caption{\label{CPasym} Leptonic CP asymmetry generated in $\Phi\to L\bar\nu$ decay at $\mathcal{O}(Y_\nu^2 Y_\ell^2)$, where the left and the right diagram represent the tree-level and the one-loop contribution, respectively.}
\end{figure}

We now start to consider the leptonic CP asymmetry generated in $\Phi \to L \bar\nu$ decay. Since $Y_\nu \ll Y_\ell$ is a generic condition for realizing the DN, we shall not consider the CP asymmetry generated at $\mathcal{O}(Y_\nu^4)$, which is the case in seesaw-based leptogenesis~\cite{Buchmuller:2004nz}. Instead, we shall determine the CP asymmetry generated at $\mathcal{O}(Y_\nu^2 Y_\ell^2)$. At this order, the absorptive part of the decay amplitude could arise from the self-energy diagrams in the lepton-doublet sector, as well as from the vertex diagrams if $\Phi$ also couples to the right-handed charged leptons. Here, let us concentrate on the former. Note that the contribution from the vertex diagrams is found to be of similar size as that from the self-energy diagrams in the SM Higgs case, and is even absent in the neutrinophilic 2HDM, as will be detailed in section~\ref{sec:4}. Then, the CP asymmetry may arise from the interference between the tree and the one-loop diagrams shown in figure~\ref{CPasym}. At zero-temperature regime, $T=0$, there is no on-shell cut for an electroweak scalar running in the loop. At high-temperature regime, however, due to the thermal bath corrections, the propagators can be on shell, producing therefore a nonzero absorptive part in the amplitude~\cite{Giudice:2003jh,Hambye:2016sby}.
 
The amplitude for $\Phi \to L \bar\nu$ decay can be defined as $i \mathcal{M}\equiv c_0 I_0+c_1 I_1$, where the coupling constants have been factored out into $c_{0,1}$, while all the other factors are contained in $I_{0,1}$, with the subscripts $0$ and $1$ referring respectively to the contributions from the tree and one-loop diagrams shown in figure~\ref{CPasym}. The thermal leptonic CP asymmetry is then given by
\begin{align}\label{CP definition}
\epsilon_D\equiv \frac{\Gamma(\Phi\to L \bar\nu)-\Gamma(\bar\Phi\to \bar L \nu)}{\Gamma(\Phi\to L \bar\nu)+\Gamma(\bar\Phi\to \bar L \nu)}
\simeq -2\frac{\text{Im}(c_0^* c_1)}{\vert c_0\vert^2} \frac{\text{Im}(I_0^*I_1)}{\vert I_0\vert^2},
\end{align}
where the second equation is obtained in the rest frame of the Higgs doublet $\Phi$. Since in the DN mechanism, the final baryon-number asymmetry can be determined from the lepton-number asymmetry stored in the right-handed neutrino sector via eq.~\eqref{sphaleron relation}, the information about the evolution of lepton-number asymmetries among the three lepton-doublet flavors is not necessary. Given that the three $\nu_R$ species are out of equilibrium, we should sum over all the interactions that contribute to the total $\nu_R$ CP asymmetry. This is implemented by summing over all the decay flavor channels of the Higgs doublet $\Phi$ in eq.~\eqref{CP definition}, as was done in the conventional DN framework. With the flavor indices being specified in figure~\ref{CPasym}, we have $c_0=Y_{\nu,ij}$ and $c_1=Y_{\nu, kj} Y^*_{\ell,kl} Y_{\ell, il}$. It can be seen that diagonal $Y_\nu$ or $Y_\ell$ would lead to $\text{Im}(c_0^* c_1)=0$. Taking into account the thermal masses and neglecting the small neutrino masses, we obtain the tree-level amplitude squared as
\begin{align}
\vert I_0\vert^2=M_\Phi^2(T)-m_{L_i}^2(T).
\end{align}
Here we should mention that the thermal corrections to fermions would modify the Dirac equation for a spinor $\psi$ to $[(1+a)\slashed p+b\slashed u] \psi=0$~\cite{Weldon:1982bn}, where $u$ is the four-velocity of the thermal bath, and $a,b$ are temperature-dependent functions (see \textit{e.g.}, ref.~\cite{Giudice:2003jh}). The functions $a,b$ would also modify the fermion propagators and hence the dispersion relations, making the expressions for spin summation and propagator poles quite lengthy and involved. Nevertheless, as illustrated in ref.~\cite{Giudice:2003jh}, to a good approximation, both the dispersion relations and the modified Dirac equations can be simplified by replacing the vacuum mass with the thermal one, \textit{i.e.}, $p^2\simeq m^2(T)$. We shall confine ourselves to adopt such an approximation in the subsequent calculations.

\subsubsection{Thermal effects: time-ordered and retarded/advanced cuts}
\label{sec:TO&RAcuts}

To calculate the leptonic CP asymmetry arising purely from the thermal effects, one can either use the non-equilibrium QFT or the thermal field theory outlined above. However, it was noticed that in the seesaw-based leptogenesis, \textit{i.e.}, for the heavy Majorana neutrino decay $N\to H L$, the CP asymmetry calculated in the non-equilibrium QFT depends linearly on the distribution functions, while quadratically in the thermal field theory~\cite{Garny:2009rv,Garny:2009qn}. It was demonstrated later in ref.~\cite{Garny:2010nj} that such a discrepancy for this process can be resolved, if the conventional time-ordered (TO) cut~\cite{Kobes:1986za,Kobes:1990ua,Gelis:1997zv,Giudice:2003jh} is replaced by the retarded/advanced product (dubbed as retarded/advanced (RA) cut hereafter for comparison)~\cite{Kobes:1990kr,Kobes:1990ua}. Here we shall check whether the same observation can be made for a DN process. To this end, we present both the TO and RA cuts within the thermal field theory, and then discuss the leptonic CP asymmetries with both cutting schemes.

\begin{figure}[t]
	\centering	
	\includegraphics[width=0.42\textwidth]{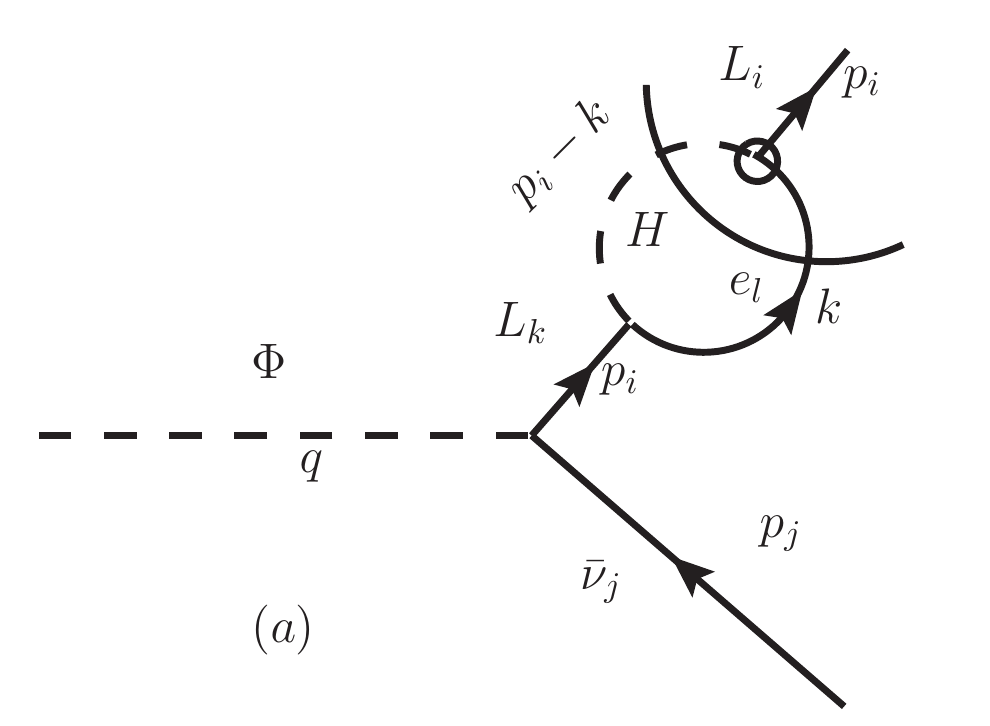} \qquad
	\includegraphics[width=0.42\textwidth]{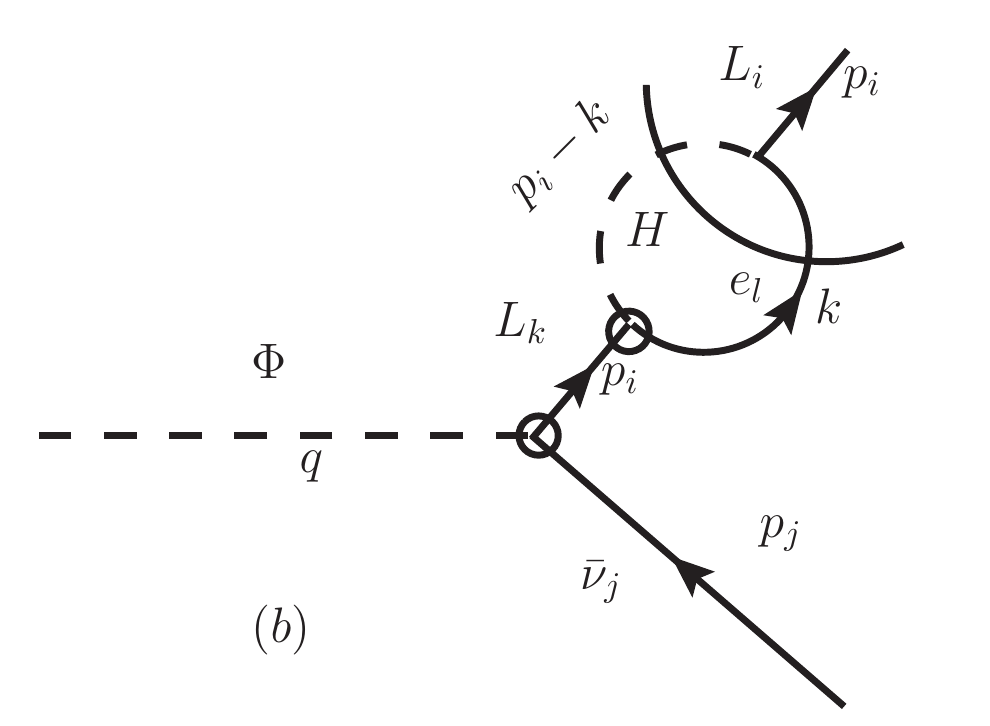}
	\caption{\label{thermalcuts} Thermal TO cuts (circlings) for producing a non-vanishing CP asymmetry in $\Phi\to L_i \bar\nu_{j}$ decay. The external thermal indices are fixed to $+$, while the internal ones are summed over $\pm$.}
\end{figure}

Following our definitions for the amplitudes $I_{0,1}$, the imaginary (absorptive) part of the product $I_0^* I_1$ in eq.~\eqref{CP definition} can be written as
\begin{align}\label{imagianry amp}
\text{Im}(I_0^* I_1)=\frac{1}{2i}I_0^* \sum_{\text{circling}}I_1.
\end{align}
With the TO-cutting scheme, there are two circling diagrams contributing to the CP asymmetry, as shown in figure~\ref{thermalcuts}. Summing the internal thermal indices over $\pm$, while fixing the external ones to $+$, we can write the corresponding amplitudes as
\begin{align}
I_1^{(a)} &= i\int\frac{d^4 k}{(2\pi)^4} \left(\bar{u}_{L_i} P_R\, u_{e_l} \right)\left(\bar{u}_{e_l} P_L u_{L_k} \right)\left(\bar{u}_{L_k} P_R v_{\nu_j} \right)\,\nonumber\\[0.1cm]
&\hspace{0.5cm} \times G^F_{++}(p_i)\,G^F_{+-}(k)\,G^B_{+-}(p_i-k),
\\[0.2cm]
I_1^{(b)} & = -i\int\frac{d^4 k}{(2\pi)^4} \left(\bar{u}_{L_i} P_R\, u_{e_l} \right)\left(\bar{u}_{e_l} P_L u_{L_k} \right)\left(\bar{u}_{L_k} P_R v_{\nu_j} \right)\,\nonumber\\[0.1cm]
&\hspace{0.5cm} \times G^F_{--}(p_i)\,G^F_{-+}(k)\,G^B_{-+}(p_i-k),
\end{align}
where the superscripts $B$ and $F$ denote the bosonic and fermionic propagators, respectively. The absorptive part for the TO cut is then determined to be
\begin{align}\label{ImIoI1}
\text{Im}(I_0^* I_1)^{ \text{TO}} &= \frac{1}{2 (2\pi)^2}\int d\omega\; \vert\boldsymbol{k}\vert^2 d \vert\boldsymbol{k}\vert\; d\cos\theta \; d\varphi \times \text{Tr}
\nonumber \\[0.1cm]
&\times \frac{1}{p_i^2-m_{L_k}^2} \times \delta[k^2-m_{e_l}^2]\times \delta[(p_i-k)^2-M_H^2]
\nonumber \\[0.1cm]
&\times \Big\{ \left[\theta(-\omega)-f_F(\vert \omega\vert)\right] \left[\theta(-(E_i-\omega))+f_B(\vert E_i-\omega\vert)\right]
\nonumber \\[0.1cm]
&\quad + \left[\theta(\omega)-f_F(\vert \omega\vert)\right] \left[\theta(E_i-\omega)+f_B(\vert E_i-\omega\vert)\right] \Big\},
\end{align}
where the four-momenta $k$ and $p_i$ are decomposed, respectively, as $k=(\omega,\boldsymbol{k})$ and $p_i=(E_i, \boldsymbol{p_i})$, while $\cos\theta\equiv \boldsymbol{p_i}\cdot \boldsymbol{k}/\vert\boldsymbol{p_i}\vert \vert\boldsymbol{k}\vert$. The trace from the spin summation is given by
\begin{align}
\text{Tr}=(k\cdot p_i) \left(4 q\cdot p_i-2m_{L_i}^2\right)-2m_{L_i}^2 (k\cdot q).
\end{align}

To perform the integration in eq.~\eqref{ImIoI1}, a convenient way is to integrate firstly over $\cos\theta$ via the Dirac delta function $\delta[(p_i-k)^2-M_H^2]$, then over $\vert\boldsymbol{k}\vert$ via $\delta[k^2-m_{e_l}^2]$, and finally over $\omega$. Due to the appearance of Heaviside step functions in eq.~\eqref{ImIoI1}, however, we must determine the sign of $\omega$ before performing the integration over $\omega$. To this end, keeping in mind that $M_{H}$ is much larger than $m_{L_i}$ and $m_{e_l}$, the presence of $\delta[(p_i-k)^2-M_H^2]$, together with $\delta[k^2-m_{e_l}^2]$, implies that
\begin{align}\label{deltamil}
\Delta m_{il}^2\equiv m_{L_i}^2+m_{e_l}^2-M_{H}^2=2p_i\cdot k=2(E_i \omega -\vert\boldsymbol{p_i}\vert \vert\boldsymbol{k}\vert \cos\theta)<0. 
\end{align}
With $-1\leqslant\cos\theta \leqslant 1$, it can then be found that  
\begin{align}\label{oemga sign}
\omega<0,\quad  E_i-\omega>0. 
\end{align}
As a consequence, the overall dependence of eq.~\eqref{ImIoI1} on the distribution functions is now simplified as
\begin{align}\label{quadratic dependence}
N(\omega)\equiv f_B(\vert E_i-\omega\vert)-f_F(\vert \omega\vert)-2f_F(\vert \omega \vert)f_B(\vert E_i-\omega\vert).
\end{align}
The final integration region of $\omega$ is determined by
\begin{align}
-1\leqslant\frac{\Delta m_{il}^2-2E_i \omega}{-2\vert\boldsymbol{p_i}\vert \vert\boldsymbol{k}\vert}\leqslant 1,
\end{align}
where $\vert\boldsymbol{k}\vert $ takes the approximate dispersion relation, $\vert\boldsymbol{k}\vert=\sqrt{\omega^2-m_{e_l}^2}$, resulting therefore in $\omega_{min}\leqslant\omega\leqslant\omega_{max}$, with
\begin{align}
\omega_{min}&=\frac{1}{4\,M_{\Phi}\,m_{L_i}^2}\left[\Delta m_{il}^2\, (M_{\Phi}^2+m_{L_i}^2)- (M_\Phi^2-m_{L_i}^2)\,\sqrt{\Delta m_{il}^4-4m_{e_l}^2 m_{L_i}^2}\,\right],
\nonumber \\[0.2cm]
\omega_{max}&=\frac{1}{4\,M_{\Phi}\,m_{L_i}^2}\left[\Delta m_{il}^2\, (M_{\Phi}^2+m_{L_i}^2)+ (M_\Phi^2-m_{L_i}^2)\,\sqrt{\Delta m_{il}^4-4m_{e_l}^2 m_{L_i}^2}\,\right],
\end{align}
in the limit of vanishing neutrino masses.  Our final expression of the CP asymmetry is then given by
\begin{align}\label{CP result}
\epsilon_D = \frac{-2\sum\limits_{i\neq k}\text{Im}[(Y_\nu Y_\nu^\dagger)_{ki}(Y_{\ell,il}Y^\dagger_{\ell,lk})]}{\sum\limits_j (Y_\nu Y^\dagger_\nu)_{jj}(M_\Phi^2-m_{L_j}^2)}
\, \mathcal{F}(M_\Phi^2,m_{L_i}^2,m_{L_k}^2,m_{e_l}^2),
\end{align}
with the scalar function defined by
\begin{align}\label{scalar fun}
\mathcal{F}(M_\Phi^2,m_{L_i}^2,m_{L_k}^2,m_{e_l}^2)&=\frac{1}{8\pi}\frac{M_\Phi^2}{(M_\Phi^2-m_{L_i}^2)(m_{L_i}^2-m_{L_k}^2)}
\nonumber \\[0.1cm]
&\times \int_{\omega_{min}}^{\omega_{max}} d\omega \left(\Delta m_{il}^2 M_\Phi-2m_{L_i}^2\omega\right) N(\omega).
\end{align}
It should be emphasized here that the dependence of $\mathcal{F}$ on the index $l$ comes from the charged-lepton Yukawa coupling contribution to the thermal lepton mass $m_{e_l}^2$ present in $\Delta m_{il}^2$ (see eq.~\eqref{deltamil}). As $\Delta m_{il}^2$ is dominated by contributions from the gauge and top-quark Yukawa couplings (the thermal masses will be discussed later in section~\ref{sec:4}), it can be inferred that the $l$ dependence is very weak, rendering therefore the CP asymmetry to carry an imaginary piece, $\text{Im}[(Y_\nu Y_\nu^\dagger)_{ki}(Y_{\ell}Y^\dagger_{\ell})_{ik}]$, as mentioned already in the Introduction. Furthermore, the $i$ dependence coming from $M_\Phi^2-m_{L_i}^2$ is also overwhelmed by contributions from the gauge couplings, as well as the possibly sizable scalar potential parameters and quark Yukawa couplings. However, as the dominant contributions from the gauge couplings are canceled out in $m_{L_i}^2-m_{L_k}^2$, the $i,k$ dependence coming from $m_{L_i}^2-m_{L_k}^2$ cannot be neglected. Thus, the CP asymmetry given by eq.~\eqref{CP result} displays a nontrivial dependence on the indices $i,k$. 

As can be seen from eq.~\eqref{quadratic dependence}, using the TO cut, we have obtained a quadratic dependence of the CP asymmetry on the distribution functions. Such a quadratic dependence was also derived for the thermal Higgs decay $H\to NL$ in ref.~\cite{Giudice:2003jh}. Within the non-equilibrium QFT framework~\cite{Frossard:2012pc}, however, the dependence was found to be linear in $f_H+f_L$ for the same decay when a retarded self-energy cut is adopted~\cite{Hambye:2016sby}. Following the argument made in ref.~\cite{Garny:2010nj}, we now turn to use the RA cut to determine the absorptive part, and check whether such a linear dependence can be reproduced in our case. With our convention, the imaginary amplitude is given by
\begin{align}
\text{Im}(I_0^*I_1)^{\text{R/A}}=\mp \frac{1}{2i}I_0^* \sum^{RA\;\text{cut}}_{\text{circling}}I_1,
\end{align}
where $\mp$ correspond to the results obtained with the retarded/advanced cut, respectively. In this context, only the circling diagram shown in figure~\ref{thermalcuts}(a) contributes, leading therefore to 
\begin{align}\label{RAcut_ImI0I1}
\text{Im}(I_0^*I_1)^{\text{R/A}}&=\mp \frac{1}{2i}\int \frac{d^4 k}{(2\pi)^4}  \left(\bar{u}_{L_i}  P_R v_{\nu_j} \right)^* \left(\bar{u}_{L_i}  P_R u_{e_l} \right)\left(\bar{u}_{e_l}  P_L u_{L_k} \right)\left(\bar{u}_{L_k}   P_R v_{\nu_j} \right) \nonumber \\[0.1cm]
&\hspace{0.5cm} \times \left[D_{L_k}(p_i)D^+_{e_l}(k)D_{H}^+(p_i-k)-D_{L_k}(p_i)D^-_{e_l}(k)D_{H}^-(p_i-k)\right] \nonumber \\[0.2cm]
&=\pm\frac{1}{2(2\pi)^2}\int d^4k \frac{1}{m_{L_i}^2-m_{L_k}^2}\delta[k^2-m_{e_l}^2]\delta[(p_i-k)^2-M_H^2] \times \text{Tr} \nonumber \\[0.1cm]
&\hspace{0.5cm} \times \left[f_F(-\omega)+f_B(E_i-\omega)\right],
\end{align}
where the thermal propagators with the RA cut are given, respectively, by
\begin{align}
 D(p)=G_{++}(p), ~ D^-(p)=G_{+-}(p), ~ D^+(p)=G_{-+}(p),
\end{align}
and eq.~\eqref{oemga sign} has been used. It can be clearly seen from eq.~\eqref{RAcut_ImI0I1} that the RA-cutting scheme does lead to a linear dependence on the distribution function $f_B+f_F$. At the same time, the imaginary amplitude, $\text{Im}(I_0^*I_1)^{\text{TO}}$, obtained with the TO-cutting scheme can be reconciled to match the retarded amplitude, $\text{Im}(I_0^*I_1)^{\text{R}}$, via the following replacement for the distribution functions:
\begin{align}\label{distribution dependence}
f_B-f_F-2f_B f_F\to f_B+f_F. 
\end{align}

We have thus shown explicitly that, when the RA-cutting scheme is used in the thermal field theory, the leptonic CP asymmetry for a DN process also displays a linear dependence on the distribution functions, which is the same as what would be obtained in the non-thermal QFT framework. Such a consistency implies that, if the leptonic CP asymmetry is calculated in the thermal field theory, a retarded amplitude would be more reliable. Note also that the RA-cutting scheme has been proved to be quite efficient for carrying actual calculations, and to have very close connections with the imaginary-time formalism~\cite{Gelis:1997zv}. In particular, in calculating the imaginary part of a Green's function at finite temperature, it is also the retarded amplitude calculated in the real-time formalism that can coincide with the one calculated in the imaginary-time formalism~\cite{Das1997}. Thus, we shall use exclusively the retarded amplitude in the subsequent analyses.
	
\subsubsection{Simplified Boltzmann equation: freeze-in evolution}

With the leptonic CP asymmetry in hand, we now proceed to determine the evolution of the lepton-number asymmetry. The generic Boltzmann equation for species $X$ participating in the process $A+B\rightleftarrows C+X$ reads 
\begin{align}\label{Boltzmann eq}
\dot{n}_X+3H n_X & = \int d\Pi_X\, d\Pi_A\, d\Pi_B\, d\Pi_C\, (2\pi)^4\,\delta^{(4)}(p_A+p_B-p_C-p_X) \\
&\hspace{-2.0cm} \times \Big[\vert\mathcal{M}_{A+B\to C+X}\vert^2\,f_A f_B (1\pm f_C)(1\pm f_X) -\vert \mathcal{M}_{C+X\to A+B}\vert^2 f_C f_X (1\pm f_A)(1\pm f_B) \Big],\nonumber
\end{align}
where the phase-space factor is given by $d\Pi_i=d^3p_i/((2\pi)^32E_i)$. 
The Dirac delta function $\delta^{(4)}(p_A+p_B-p_C-p_X)$ enforces the four-momentum conservation in the collisions. The amplitude squared is obtained by summing over the initial- and final-state spins but without average. The factors $1\pm f_i$ correspond to the Bose enhancement and the Pauli blocking effect, respectively. The Hubble parameter at radiation-dominated flat Universe is given by $H=1.66\sqrt{g^\rho_\ast}\,T^2/M_{Pl}$, where $g^\rho_*$ denotes the relativistic d.o.f at temperature $T$, and $M_{Pl}=1.2\times 10^{19}$~GeV is the Planck mass. 

In the early Universe, right-handed neutrinos are produced only through feeble Yukawa interactions. Thus, the production is essentially out-of-equilibrium and the particle number density is therefore negligibly small. Such a freeze-in production mechanism~\cite{Hall:2009bx} effectively prevents large washout effects in the neutrino number density (as well as the lepton-number asymmetry generated therein) from inverse decay and annihilation scattering. In this context, the Boltzmann equation for the lepton-number asymmetry generated in $\Phi\to L \bar \nu$ decay can be much simplified by neglecting the inverse decay and annihilation scattering, because these processes are proportional to the negligible particle-number density. As a consequence, the lepton-number asymmetry can be accumulated as the right-handed neutrinos are produced, and then converted to the baryon-number asymmetry via rapid sphaleron transitions. 

Based on the above observation, the Boltzmann equation for the right-handed neutrinos can be, therefore, simplified as
\begin{align}
\dot{n}_{\nu}+3H n_{\nu} =\int d\Pi_{\Phi} f^{eq}_\Phi \int d\Pi_{\nu} d\Pi_L\,  (2\pi)^4\delta^{(4)}(p_\Phi-p_{\nu}-p_L)\,\vert\mathcal{M}(\bar\Phi\to \bar L \nu)\vert^2,
\end{align}
where, as an approximation, we have set the quantum statistic factors $1\pm f\approx1$. Together with the definition of the CP asymmetry $\epsilon_D$ (see eq.~\eqref{CP definition}), the evolution of the lepton-number asymmetry $n_{\Delta \nu}\equiv n_{\nu}-n_{\bar \nu}$ can be written as
\begin{align}\label{lepton-asymmetry Boltzmann }
\dot{n}_{\Delta \nu}+3Hn_{\Delta \nu}&\approx \int d\Pi_\Phi f_{\Phi}^{eq}\, 2\,g_{\Phi} \,M_{\Phi}\,\left[\Gamma(\bar\Phi\to \bar L \nu)-\Gamma(\Phi \to L \bar\nu)\right] \nonumber\\
&=-\int d\Pi_{\Phi}\,f_{\Phi}^{eq}  \times 2\,g_{\Phi}\,M_{\Phi} \times 2\,\epsilon_D \times \Gamma(\Phi\to L \bar\nu),
\end{align} 
where $g_{\Phi}=2$ results from the two gauge components of the Higgs doublet $\Phi$. Note that here we have neglected the contribution from the $\Phi$ asymmetry. Since $\Phi$ is presumed to keep in equilibrium with the thermal bath during the neutrinogenesis, such an asymmetry generated non-thermally in the decay process is proportional to the neutrino Yukawa couplings, and hence negligible with respect to the symmetric portion produced from the thermal interactions. In addition, since the thermal particle $\Phi$ is assumed to take a vacuum mass around $\mathcal{O}(10^2)$~GeV, we can use, as a good approximation, the Maxwell-Boltzmann distribution $f^{eq}_{\Phi}=f^{eq}_{\bar \Phi}=e^{-E/T}$ for the phase-space integration in the above equations. At the sphaleron-active epoch, the relativistic d.o.f $g_*^s$ present in the entropy density, $s=T^3 g_*^s 2\pi^2/45$, can be treated as a constant, and thus we can use the yield definition $Y=n/s$, with $\dot{s}=-3Hs$ and $\dot{T}=-H T$, to rewrite the above equation as
\begin{align}\label{lepton-asymmetry Boltzmann2}
\frac{dY_{\Delta \nu}}{dT}=\frac{1}{sH}\,\frac{1}{\pi^2}\,g_{\Phi}\,\epsilon_D\,M_{\Phi}^2\,  K_1(M_{\Phi}/T)\,\Gamma(\Phi\to L\bar\nu),
\end{align} 
where $K_1$ denotes the first modified Bessel function of the second kind. In the rest frame of $\Phi$, the decay width is given by
\begin{align}\label{decay rate}
\Gamma(\Phi \to L \bar\nu)&=\sum\limits_{i=e,\mu,\tau}\frac{1}{8\pi}\,(Y_\nu Y_\nu^\dagger)_{ii}\,\frac{\vert \boldsymbol{p_i}\vert}{M_{\Phi}^2}\,(M_{\Phi}^2-m_{Li}^2),
\end{align}
where $\vert\boldsymbol{p_i}\vert=(M_{\Phi}^2-m_{Li}^2)/(2M_{\Phi})$ denotes the magnitude of the three-momentum of the two final-state particles, and the neutrino masses have been neglected here.

Up to now, we have obtained a nonzero kinetic phase contained in $\text{Im}(I_0^*I_1)$ within the framework of thermal field theory. To generate a non-vanishing lepton-number asymmetry, however, non-diagonal Yukawa couplings $Y_{\ell,\nu}$ are also required. In the next section, we turn to exploit the nontrivial Yukawa structures that can generate a nonzero coupling phase contained in $\text{Im}(c_0^* c_1)$ and, at the same time, produce compatible neutrino oscillation data.

\section{Non-diagonal Yukawa textures in the lepton sector}
\label{sec:3}

When the lepton-number asymmetry is generated via the freeze-in production of right-handed neutrinos, the washout effects from inverse decay and annihilation scattering are negligible. Therefore, the flavor effects encoded in these washout processes, \textit{e.g.}, $\Phi \bar\nu_{i}\rightleftharpoons \bar \Phi \nu_{j}$, would not play a significant role in the Boltzmann evolution. This is also true for the flavor effects from the charged-lepton Yukawa sector, as is the case in the conventional DN framework. These observations are due to the fact that the baryon-number asymmetry is determined from eq.~\eqref{sphaleron relation} with a summation over the indistinguishable $\nu_R$ species, but without the need of specifying the detailed evolution of lepton-number asymmetry among the three lepton-doublet flavors.

However, as mentioned already below eq.~\eqref{scalar fun}, the dependence of the scalar function $\mathcal{F}$ on the indices $i,k$ from the lepton-doublet propagator cannot be neglected. In fact, such a dependence is especially crucial for producing a non-vanishing CP asymmetry $\epsilon_D$, because otherwise the imaginary coupling sector would vanish, \textit{i.e.}, $\text{Im}\big[\text{tr}(Y_\nu Y^\dagger_\nu\,Y_\ell Y^\dagger_\ell)\big]=0$. Furthermore, a nonzero $\epsilon_D$ also requires the Yukawa matrices $Y_{\ell,\nu}$ to be non-diagonal. Therefore, even after summing over the lepton flavors, the CP asymmetry is still texture dependent and relates intimately with the flavor puzzle, which is a generic feature of leptogenesis~\cite{Branco:2002kt}. 

There is thus far a plethora of model buildings aimed at deciphering the non-trivial Yukawa structures that can also lead to compatible flavor mixings observed at experiments (see, \textit{e.g.}, refs.~\cite{Altarelli:2010gt,Xing:2019vks,Feruglio:2019ktm} for comprehensive reviews). However, since we focus on exploiting whether the lepton mixing from non-diagonal Yukawa matrices can produce compatible neutrino oscillation observables and, at the same time, indicate a successful freeze-in DN mechanism, we shall refrain from performing explicit and model-dependent buildings of the Yukawa structures. Rather, we shall postulate that the observed mixing pattern of the PMNS matrix is induced by the well-known TB mixing with a minimal correction from the charged-lepton or neutrino sector~\cite{Albright:2008rp}. The popular TB mixing pattern has a mass-independent form~\cite{Harrison:2002er}:
\begin{align}
V_{TB} =\left(
\begin{array}{ccc}
 \sqrt{\frac{2}{3}}\,\,  &  \frac{1}{\sqrt{3}}\,\,   &                   0 \\[0.2cm]
-\frac{1}{\sqrt{6}}\,\,  & 	\frac{1}{\sqrt{3}}\,\,   &  \frac{1}{\sqrt{2}} \\[0.2cm]
-\frac{1}{\sqrt{6}}\,\,  & 	\frac{1}{\sqrt{3}}\,\,   & -\frac{1}{\sqrt{2}} \\
\end{array}
\right).
\end{align}
Although the pure TB mixing pattern is already excluded by the observed nonzero reactor angle $\theta_{13}$ (see \textit{e.g.}, refs.~\cite{Esteban:2016qun,deSalas:2017kay,Esteban:2018azc,Capozzi:2018ubv,deSalas:2020pgw,Esteban:2020cvm} for recent global analyses), it has been pointed out that, with a minimal correction from the charged-lepton or neutrino sector, this mixing pattern can readily produce compatible neutrino oscillation data, while retaining its predictability and testability of the relations among the mixing angles~\cite{Albright:2008rp,He:2011gb}. 

As demonstrated explicitly in ref.~\cite{He:2011gb}, there exist four possible minimal corrections to the TB mixing pattern that are still compatible with the current PMNS data at $3\sigma$ level. According to which column or row of the TB matrix is invariant under the minimal corrections, the modified patterns can be classified as $\text{TM}_i$ (invariance of the $i$-th column) and $\text{TM}^i$ (invariance of the $i$-th row)~\cite{Albright:2008rp}. Thus, on account of the observations made in refs.~\cite{Albright:2008rp,He:2011gb}, we have
\begin{align}\label{four patterns}
\text{TM}_1:\;\;U&=V_{TB}R_{23},\qquad
\text{TM}_2:\;\;U=V_{TB}R_{13},
\nonumber \\[0.2cm]
\text{TM}^2:\;\;U&=R_{13}V_{TB},\qquad
\text{TM}^3:\;\;U=R_{12}V_{TB},
\end{align}
with the unitary Euler rotation matrices given, respectively, by
\begin{align}
R_{12}(\theta)&=\left(
\begin{array}{ccc}
\cos\theta\; & \sin\theta e^{i\varphi}\; & 0\\[0.1cm]
-\sin\theta e^{-i\varphi }\; & 	\cos\theta\; & 0\\[0.1cm]
0\; & 0\; & 1\\
\end{array}
\right),\qquad
R_{13}(\theta)=\left(
\begin{array}{ccc}
\cos\theta\; & 0\;    & \sin\theta e^{i\varphi }\\[0.1cm]
0\; & 1\;	& 0\\[0.1cm]
-\sin\theta e^{-i\varphi}\; &   0\;  & 	\cos\theta \\
\end{array}
\right),
\nonumber \\[0.2cm] 
& \hspace{2.3cm}  R_{23}(\theta )=\left(
\begin{array}{ccc}
1\;&                   0\;           &                 0\\[0.1cm]
0\;&  \cos\theta\;                   &  \sin\theta e^{i\varphi}\\[0.1cm]
0\;& -\sin\theta e^{-i\varphi }\;    &  \cos\theta\\
\end{array}
\right),
\end{align}
where $0\leq\theta\leq\pi$ and $0\leq\varphi<2\pi$. With the convention $M_f=V_f^{L\dagger} \hat{M}_f V_f^R$, where $M_f$ and $\hat{M}_f$ represent the mass matrices before and after the diagonalization, the flavor (primed) and the mass (un-primed) eigenstates are transformed to each other via the relations $f_{L(R)}^\prime =V_f^{L(R)\dagger} f_{L(R)}$, and the PMNS matrix is given by $U=V_\ell^L V_\nu ^{L \dagger}$. Eq.~\eqref{four patterns} indicates then that the charged-lepton and neutrino mixing matrices would be given, respectively, by $V_\ell^L=V_{TB}$ and $V_\nu^L=R_{23}^\dagger, R_{13}^\dagger$ in patterns $\text{TM}_{1,2}$, while by $V_\ell^L=R_{13}, R_{12}$ and $V_\nu^L=V_{TB}^\dagger$ in patterns $\text{TM}^{2,3}$. The product of the Yukawa matrices can also be rewritten in terms of the mixing matrix and the mass spectrum as
\begin{align}
Y_f Y_f^\dagger =\frac{2}{v_f^2}\,V_f^{L \dagger}\,\hat{M}_f^2\,V^L_f,
\end{align}
where $v_f$ denote the nonzero VEVs developed by the Higgs doublets responsible for generating the charged-lepton and neutrino masses, respectively. Given that the leptonic CP asymmetry $\epsilon_D$ is approximately proportional to $\text{Im}[(Y_\nu Y_\nu^\dagger)_{ki}(Y_{\ell}Y^\dagger_{\ell})_{ik}]$ (see eq.~\eqref{CP result}), it can be seen that, after fixing the kinematics, $\epsilon_D$ will depend on the two mixing parameters, $(\theta, \varphi)$, both of which are also directly responsible for producing the current neutrino oscillation data.

Besides the requirement that eq.~\eqref{four patterns} should produce the observed moduli of the PMNS matrix, $\vert U \vert$, a basis-independent and rephasing-invariant measure of the low-energy CP violation, defined as~\cite{Jarlskog:1985ht}
\begin{align}
\mathcal{J}\sum\limits_\gamma \epsilon_{\alpha \beta \gamma} \sum\limits_k \epsilon_{ijk}=\text{Im}[U_{\alpha i}U^*_{\alpha j}U^*_{\beta i}U_{\beta j}],
\end{align}
is also crucial to exploit how successfully the freeze-in DN can be inferred, particularly, from the sign of $\mathcal{J}$. In the standard convention of the PMNS matrix~\cite{Tanabashi:2018oca}, the Jarlskog invariant $\mathcal{J}$ is given by
\begin{align}
\mathcal{J}=\frac{1}{8}\cos\theta_{13}\sin(2\theta_{12})\sin(2\theta_{23}) \sin(2\theta_{13})\sin\delta.
\end{align}
Recently, a global fit of neutrino oscillation parameters has obtained a strong preference for values of the Dirac CP phase $\delta$ in the range $[\pi,2\pi]$~\cite{deSalas:2017kay,deSalas:2020pgw}. Together with the measured mixing angles, this implies a negative CP measure $\mathcal{J}<0$. In particular, the best-fit value favors $\delta\simeq 3\pi/2$~\cite{Esteban:2016qun,Esteban:2020cvm}, leading to (at $3\sigma$ level)
\begin{align}
\mathcal{J}^{max}_{CP}=-\left(0.0329^{+0.0021}_{-0.0024}\right),
\end{align}
where the uncertainties come from the determination of the mixing angles.

Corresponding to the four mixing patterns specified by eq.~\eqref{four patterns}, the Jarlskog invariant $\mathcal{J}$ is given, respectively, by
\begin{align}\label{Jarlskog}
\text{TM}_1&: \mathcal{J}=-\frac{\sin(2\theta)\sin\varphi}{6\sqrt{6}}, &\text{TM}_2&: \mathcal{J}=-\frac{\sin(2\theta)\sin\varphi}{6\sqrt{3}},
\nonumber \\[0.5mm]
\text{TM}^2&: \mathcal{J}=\frac{\sin(2\theta)\sin\varphi}{12}, &\text{TM}^3&: \mathcal{J}=-\frac{\sin(2\theta)\sin\varphi}{12}.
\end{align}
If the maximal CP violation, $\mathcal{J}=\mathcal{J}_{CP}^{max}$, is assumed, and the values of the mixing angles, $(\theta,\varphi)$, are taken to produce the $3\sigma$ ranges of the PMNS matrix moduli $\vert U\vert$, we can then establish whether such a low-energy maximal CP violation can prompt a successful DN. To this end, we need firstly specify the decaying particle for generating the leptonic CP asymmetry, which will be explored at length in the next section.

\section{Thermal scalar implementation}
\label{sec:4}

As an alternative to most DN applications in which the lepton-number asymmetry is generated by non-thermal heavy particle decays~\cite{Dick:1999je,Murayama:2002je,Cerdeno:2006ha,Gu:2007mc,Bechinger:2009qk,Narendra:2017uxl}, we have considered the case where the asymmetry is accumulated via the freeze-in production of right-handed neutrinos from thermal scalar decay. In this section, we shall specify the minimal Higgs doublets for implementing the freeze-in DN described in section~\ref{sec:2}, and determine the baryon-number asymmetry using the RA-cutting scheme.

\subsection{SM Higgs case}

If the DN were realized by the SM Higgs, the neutrino mass and BAU would then be simultaneously addressed by simply adding the missing neutrino Yukawa interactions (eq.~\eqref{Yukawa}) to the SM. The only price to pay is to accept the non-aesthetic, feeble neutrino Yukawa couplings, which are of $\mathcal{O}(10^{-14})$ for $\mathcal{O}(10^{-2})$~eV neutrino masses. 

Since the sphaleron-active epoch, $10^2~\text{GeV}<T<10^{12}~\text{GeV}$, is considered, we shall use the thermal masses of the SM Higgs and leptons that are given by~\cite{Weldon:1982bn,Giudice:2003jh}
\begin{align}
M_H^2 &\simeq \left(\frac{3}{16}g_2^2+\frac{1}{16}g_1^2+\frac{1}{4}y_t^2+\frac{1}{4}\lambda\right)T^2,  \label{thermal Higgs masses}
\\[0.2cm]
m_{L_i}^2 &= \left(\frac{3}{32}g_2^2+\frac{1}{32}g_1^2+\frac{1}{16}(Y_{\ell}Y^\dagger_{\ell})_{ii}\right)T^2,  \label{thermal L-doublet masses}
\\[0.2cm]
m_{e_l}^2 &= \left(\frac{1}{8}g_1^2+\frac{1}{8}(Y^\dagger_{\ell}Y_{\ell})_{ll}\right)T^2,
\label{thermal L-singlet masses}
\end{align}
where $g_{2}$~($g_{1}$) is the $SU(2)_L$~($U(1)_Y$) gauge coupling, while $\lambda$ is the SM Higgs potential parameter and satisfies the tadpole equation, $m_h^2=\lambda v^2$, with $h$ being the physical SM Higgs boson. Here we have only kept the dominant top-Yukawa contribution to $M_{H}^2$ by assuming a diagonal Yukawa matrix in the up-type quark sector. A general charged-lepton Yukawa matrix is, however, retained for the thermal masses of leptons, because a non-diagonal $Y_\ell$ is crucial for generating a non-vanishing lepton-number asymmetry. While the renormalization-group running of the coupling constants should in principle be taken into account at a scale $\mu\simeq 2\pi T$~\cite{Giudice:2003jh}, which would prompt additional $T$-dependent sources, as a simple estimation, we shall use here the vacuum values of these coupling constants. 

Based on the analysis of the index dependence of the CP asymmetry made below eq.~\eqref{scalar fun}, we can neglect the small Yukawa contributions to $M_\Phi^2 \pm m_{L_i}^2$ and $\Delta m_{il}^2$, because they are basically overwhelmed by the contributions from gauge couplings, potential parameters, and top-quark Yukawa couplings. Under this approximation, the integration of eq.~\eqref{lepton-asymmetry Boltzmann2} over the sphaleron-active regime induces a semi-analytic expression for the baryon-number asymmetry. It is found that, for the mixing patterns $\text{TM}_{1,2}$, the baryon-number asymmetry is estimated to be
\begin{align}\label{asymmetry in SM Higgs}
\text{TM}_1&:\;\; Y_{\Delta B}^{\text{R}} \simeq  -\mathcal{O}(10^{-15}) \sin(2\theta) \sin\varphi,  \\[0.2cm]
\text{TM}_2&:\;\;Y_{\Delta B}^{\text{R}}\simeq  \mathcal{O}(10^{-15}) \sin(2\theta)  \sin\varphi,
\end{align}
where $Y_{\Delta B}^{\text{R}}$ is obtained with the input of the CP asymmetry determined in the retarded-cutting scheme. For patterns $\text{TM}^{2,3}$, the baryon-number asymmetry is found to be even smaller, with
\begin{align}
\text{TM}^2&:\;\;Y_{\Delta B}^{\text{R}} \simeq  \mathcal{O}(10^{-17}) \tan(2\theta)\sin\varphi,
\\[0.2cm]
\text{TM}^3&:\;\;Y_{\Delta B}^{\text{R}}\simeq   \mathcal{O}(10^{-17}) \tan(2\theta)\sin\varphi. \label{asymmetry in SM Higgs2}
\end{align}
To obtain the above numerical factors, we have used $g_*^\rho=g_*^s=106.75$. In addition, we have adopted a normal-ordering neutrino mass hierarchy, as suggested by the recently global analyses~\cite{Capozzi:2018ubv,Esteban:2018azc}, and neglected the lightest neutrino mass. Explicitly, the values of neutrino masses used are given by $m_{\nu_1}\simeq 0$, $m_{\nu_2}\simeq \sqrt{\Delta m_{21}^2}$, and $m_{\nu_3}\simeq \sqrt{\Delta m_{31}^2}$, with the mass-squared differences taken from ref.~\cite{Esteban:2016qun}. Note that, by saturating the cosmological upper bound on the sum of neutrino masses, $\sum_i m_i<0.12$~eV~\cite{Aghanim:2018eyx}, we would obtain an upper limit of the lightest neutrino mass with a quasi-degenerate mass spectrum, $m_1=0.0301$~eV, $m_2=0.0313$~eV and $m_3=0.0586$~eV, where the best-fit values of $\Delta m_{21}^2$ and $\Delta m_{31}^2$ are used~\cite{Esteban:2016qun}. However, the modification to the generated $Y^R_{\Delta B}$ by using such a quasi-degenerate mass spectrum is only within $1\%$. This is also expected, because the summation over quasi-degenerate neutrino masses that appears in the denominator of $Y_{\Delta \nu}$ (see eqs.~\eqref{lepton-asymmetry Boltzmann2} and \eqref{CP result}) does not induce resonant-like enhancement.

It can be seen from eqs.~\eqref{asymmetry in SM Higgs}-\eqref{asymmetry in SM Higgs2} that the dependence of $Y_{\Delta B}$ on the trigonometric functions is different between the patterns $\text{TM}_{1,2}$ and $\text{TM}^{2,3}$. This is due to the fact that an additional $\theta$ dependence appears in the thermal fermion masses for patterns $\text{TM}^{2,3}$. Compared with the observed baryon-number asymmetry of the Universe at present day~\cite{Aghanim:2018eyx}, 
\begin{align}\label{BAU today}
 Y_{\Delta B}=(8.75\pm 0.23)\times 10^{-11},
\end{align}
the amount of asymmetry induced by the minimal SM Higgs is negligible. Although we have followed here a phenomenological perspective, $Y_{\Delta B}^{\text{R}}$ given by eqs.~\eqref{asymmetry in SM Higgs}-\eqref{asymmetry in SM Higgs2} are primarily controlled by the neutrino Yukawa couplings $Y_\nu\simeq \mathcal{O}(10^{-14})$, and thus the orders of magnitude estimated therein are quite reasonable. This can also be justified by noting that, even though the neutrino Yukawa couplings may be canceled in the imaginary coupling sector, the decay rate $\Gamma(H\to L \bar\nu)$ involves the couplings at $\mathcal{O}(Y_\nu^2)$. In addition, as the SM Higgs also couples to the right-handed charged leptons, an additional contribution to the leptonic CP asymmetry can be induced by the vertex correction. It is, however, expected that such an amount of asymmetry would be similar to that generated by the wavefunction correction, as no quasi-degenerate mass spectrum could resonantly enhance the latter within the SM. Based on these observations, the SM Higgs implementation should be therefore dismissed, and we are driven to consider new scalars beyond the minimal SM. 

\subsection{Neutrinophilic two-Higgs-doublet model}
\label{sec:4.2}

A direct enhancement of the lepton-number asymmetry can be achieved by invoking a sufficiently large neutrino Yukawa coupling, while retaining the out-of-equilibrium condition. This can be realized by introducing another Higgs doublet which develops a smaller VEV. As a minimal extension of the SM, let us focus on the neutrinophilic 2HDM~\cite{Davidson:2009ha}, in which the second Higgs doublet couples neither to the quarks nor to the right-handed charged leptons. 

In such a neutrinophilic 2HDM, both the right-handed Dirac neutrinos and the new Higgs doublet possess an additional $Z_2$ parity. The model Lagrangian has, however, a softly $Z_2$-breaking scalar potential~\cite{Davidson:2009ha}
\begin{align}
V&=m_1^2 H_1^{\dagger}H_1 +m_2^2H_2^{\dagger} H_2 -(m_{12}^2H_1^{\dagger} H_2 + \text{h.c.})
+\frac{\lambda_1}{2}(H_1^{\dagger} H_1)^2+ \frac{\lambda_2}{2}(H_2^{\dagger}H_2)^2
\nonumber \\[0.2cm]
&+\lambda_3(H_1^{\dagger} H_1) (H_2^{\dagger}H_2)
+\lambda_4(H_1^{\dagger} H_2) (H_2^{\dagger} H_1) + \left[ \dfrac{\lambda_5}{2}(H_1^{\dagger} H_2)^2+ \text{h.c.} \right].
\label{Higgs potential}
\end{align}
For a real and positive soft-breaking term $m_{12}^2\ll v^2$, with $v_1^2+v_2^2=v^2=(246~\text{GeV})^2$, the tadpole equations, $\partial V/\partial H_i=0$, would induce a seesaw-like relation
\begin{align}
v_1\simeq v, \qquad v_2\simeq \frac{m_{12}^2v}{ \lambda_{345} v^2+m_{2}^2},
\end{align}
with $\lambda_{345}\equiv(\lambda_3+\lambda_4+\lambda_5)/2$. In the conventional neutrinophilic 2HDM~\cite{Davidson:2009ha} (see also some phenomenological studies of the model performed in refs.~\cite{Machado:2015sha,Bertuzzo:2015ada}), the value of $v_2$ is tuned at eV scale, to have $\mathcal{O}(1)$ neutrino Yukawa couplings. Apparently, when $Y_\nu\simeq \mathcal{O}(1)$, neutrinos would establish the L-R equilibration in the sphaleron-active epoch, and thus no net lepton-number asymmetry would be stored. Here we assume, instead, $Y_\nu\lesssim \mathcal{O}(10^{-8})$ to guarantee the out-of-equilibrium generation of the lepton-number asymmetry. Such an assumption is justified by the requirement $v_2\gtrsim \mathcal{O}(10^{-3})$~GeV, which in turn indicates that $m_{12}\gtrsim0.5$~GeV for $m_{2}\simeq \mathcal{O}(v)$.

At a temperature well above the electroweak scale, we use the following thermal mass for the second Higgs doublet~\cite{Cline:1995dg}:
\begin{align}
M_{H_2}^2\simeq \left(\frac{3}{16}g_2^2+\frac{1}{16}g_1^2+\frac{1}{4}\lambda_2+\frac{1}{6}\lambda_3+\frac{1}{12}\lambda_4\right)T^2,
\end{align}
where marginal contributions from the softly $Z_2$-breaking term and the neutrino Yukawa couplings have been neglected. The thermal masses of $H_1$ and leptons are the same as that given by eqs.~\eqref{thermal Higgs masses}-\eqref{thermal L-singlet masses}. For $M_{H_2}$ present in eq.~\eqref{lepton-asymmetry Boltzmann2}, we include also the positive mass parameter $m_2$, \textit{i.e.}, $M_{H_2}^2=m_2^2+M_{H_2}^2(T)$. For our numerical analyses, we shall assume that the effects from $\lambda_{4,5}$ are negligible. Furthermore, we shall work in the alignment limit where $H_1$ contains the SM Higgs boson. Within such a numerical setup, the Higgs mass spectrum is nearly degenerate, $M_{H^\pm}\simeq M_{H}\simeq M_{A}$, and $\lambda_1\simeq \lambda_2\simeq \lambda_3\simeq m_h^2/v^2$. For explicit expressions of the potential parameters $\lambda_{1-5}$ and the Higgs mass spectrum in the alignment limit, together with the theoretical and experimental constraints on the model parameters, the readers are referred to, \textit{e.g.}, ref.~\cite{Li:2018aov}. It can also be found from refs.~\cite{Machado:2015sha,Bertuzzo:2015ada} that such a numerical mass spectrum is phenomenologically viable.

As done in the last subsection, here the numerical integration of eq.~\eqref{lepton-asymmetry Boltzmann2} over the sphaleron-active regime also prompts semi-analytic expressions for the baryon-number asymmetry $Y_{\Delta B}$:
\begin{align}
\text{TM}_1&:\; \;Y_{\Delta B}^{\text{R}}\simeq -1.88 \times 10^{-10}\;\frac{\sin(2\theta)\sin\varphi}{(v_2/\text{GeV})^2},\label{baryon asym in pattern-1}
\\[0.2cm]
\text{TM}_2&: \;\; Y_{\Delta B}^{\text{R}}\simeq 0.68\times 10^{-10}\;\frac{\sin(2\theta)\sin\varphi}{(v_2/\text{GeV})^2},\label{baryon asym in pattern-2}
 \\[0.2cm]
\text{TM}^2&:\; \; Y_{\Delta B}^{\text{R}}\simeq 0.79\times 10^{-12}\;\frac{\tan(2\theta)\sin\varphi}{(v_2/\text{GeV})^2}, \label{baryon asym in pattern-3}
 \\[0.2cm]
\text{TM}^3&:\;\;Y_{\Delta B}^{\text{R}}\simeq 0.79 \times 10^{-12}\,\frac{\tan(2\theta) \sin\varphi}{(v_2/\text{GeV})^2},\label{baryon asym in pattern-4}
\end{align}
where $g_*^\rho=g_*^s=110.75$ are used in the 2HDM framework, and we have picked a particular mass parameter $m_2=500$~GeV. In addition, we have adopted the normal-ordering neutrino mass spectrum, $m_{\nu_1}\simeq 0$, $m_{\nu_2}\simeq \sqrt{\Delta m_{21}^2}$ and $m_{\nu_3}\simeq \sqrt{\Delta m_{31}^2}$. It is also found that, taking the quasi-degenerate mass spectrum, the generated $Y^R_{\Delta B}$ is modified only within $1\%$, which is similar to what is already observed in the SM Higgs case.

\begin{figure}[ht]
	\centering
	\includegraphics[width=0.46\linewidth]{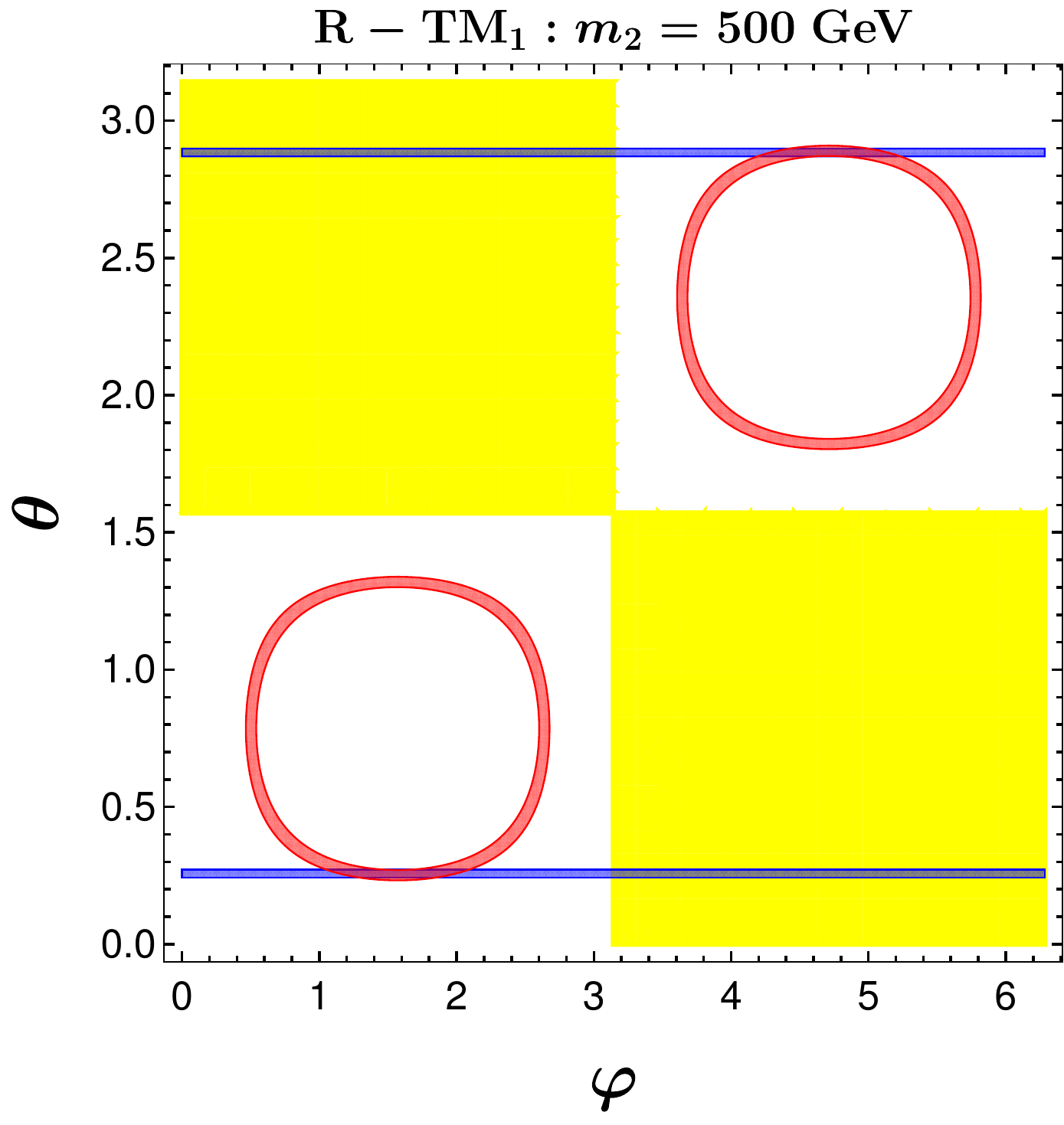}\qquad
	\includegraphics[width=0.46\linewidth]{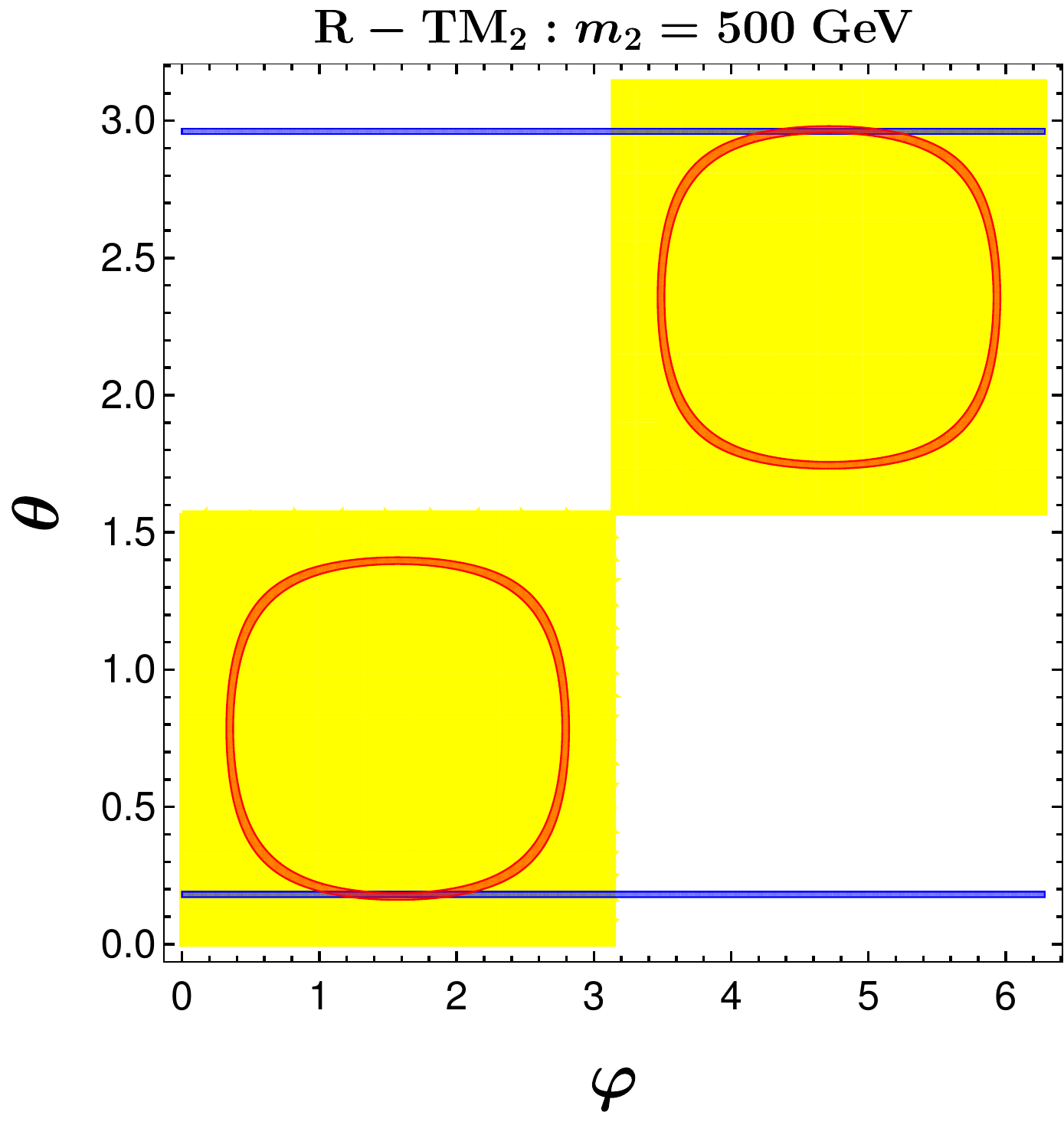}\\[0.3cm]
	\includegraphics[width=0.46\linewidth]{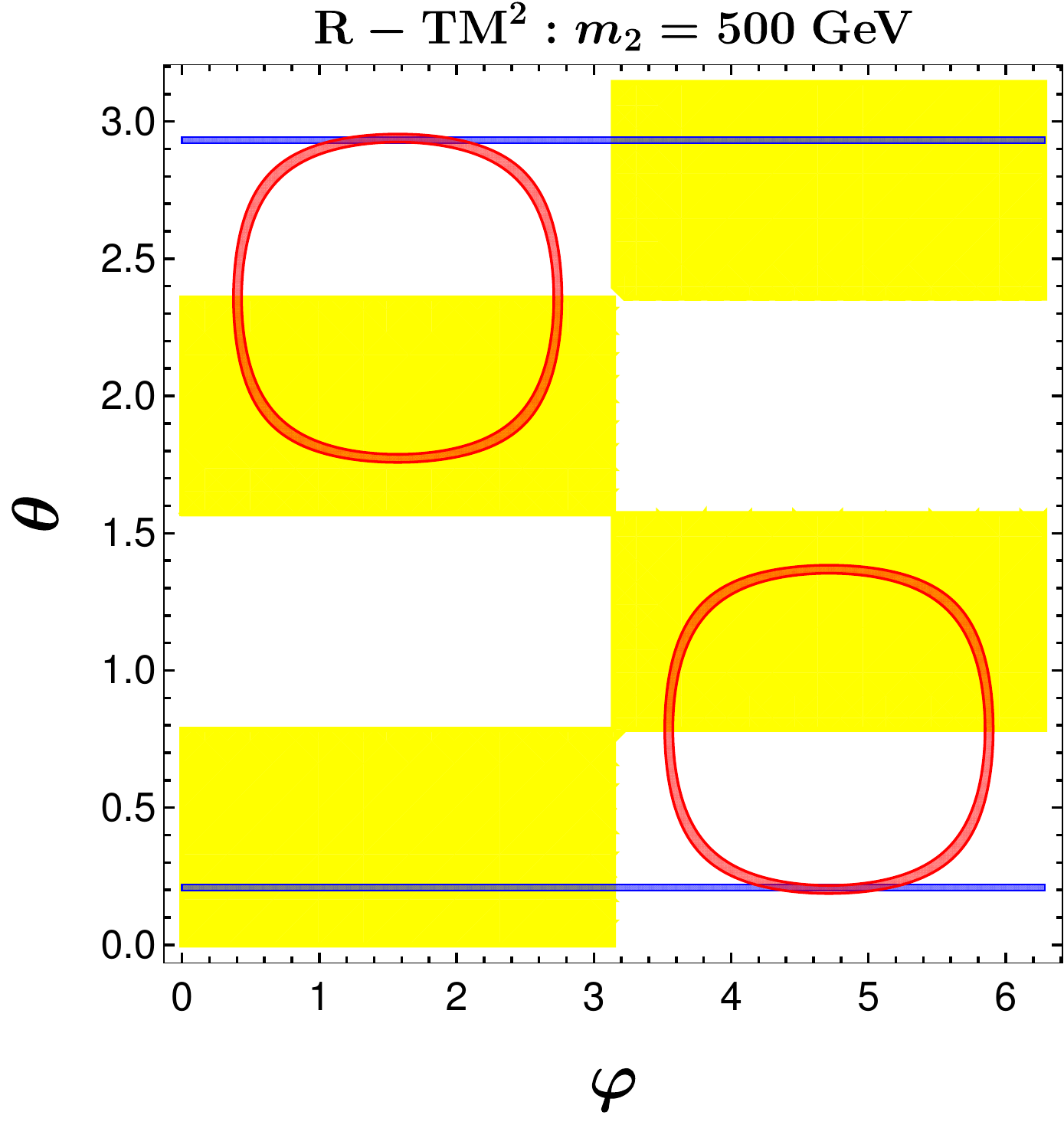}\qquad
	\includegraphics[width=0.46\linewidth]{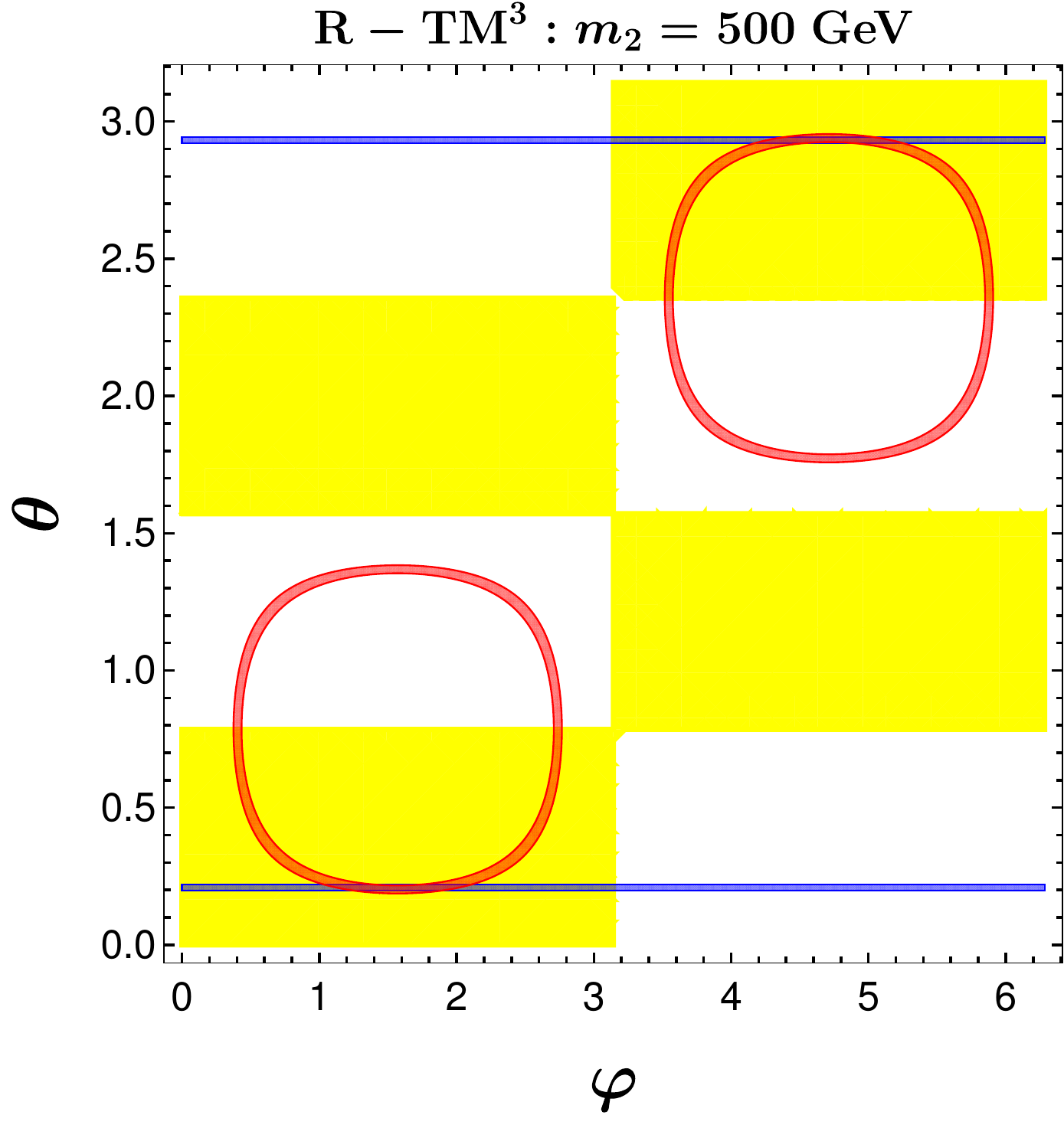} 
	\caption{\label{R-TM-Angle} Allowed regions for the two mixing parameters, $(\theta, \varphi)$, under the individual constraint from a positive baryon-number asymmetry (yellow bands), a negative CP measure $\mathcal{J}=\mathcal{J}_{CP}^{max}<0$ (red regions), as well as the PMNS matrix element $\vert U_{13}\vert$ (narrow blue bands) in neutrino oscillations~\cite{Esteban:2016qun}. Note that the yellow bands are   independent of the mass parameter $m_2$ chosen around the electroweak scale.}
\end{figure}

\begin{figure}[t]
	\centering
	\includegraphics[width=0.46\textwidth]{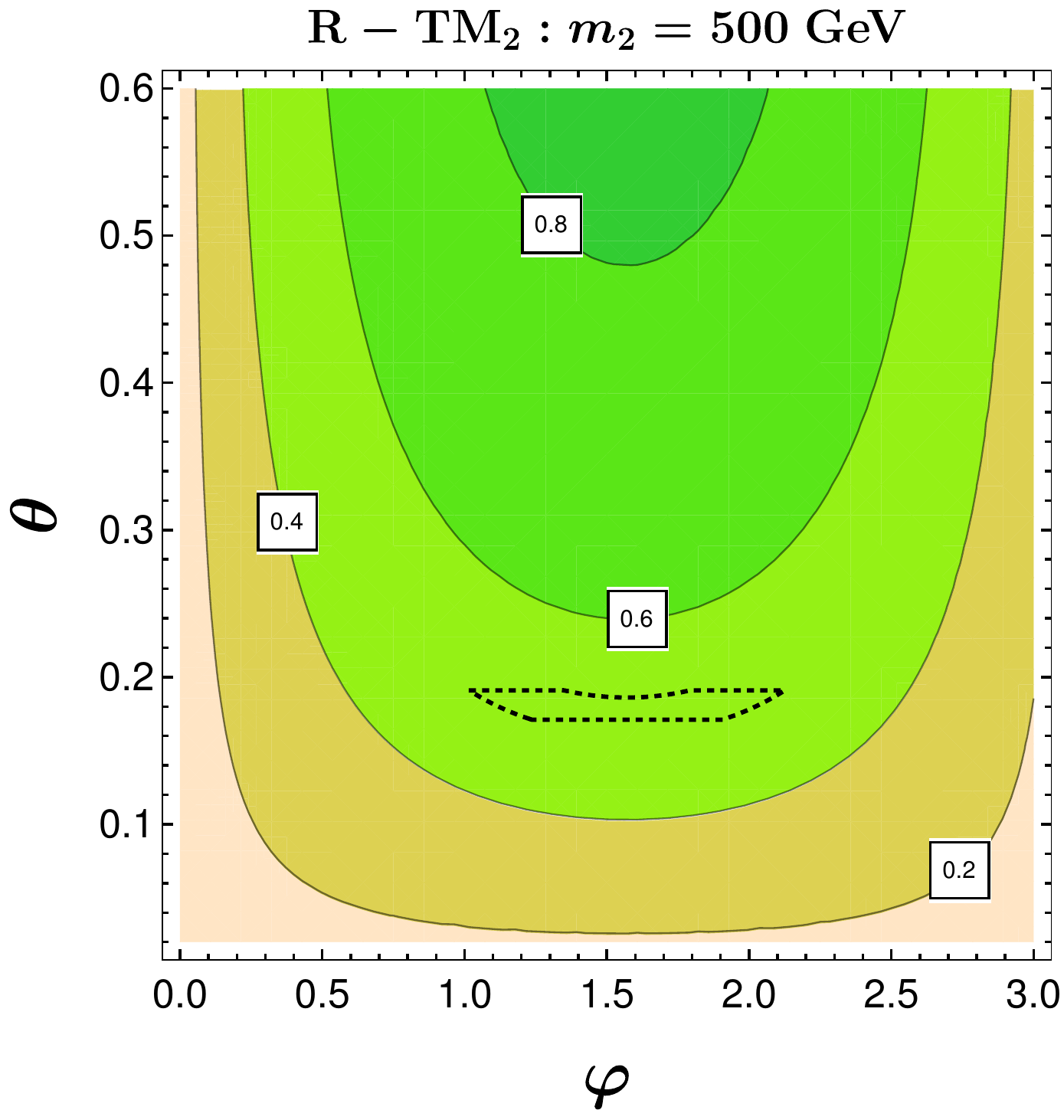}\qquad
	\includegraphics[width=0.46\textwidth]{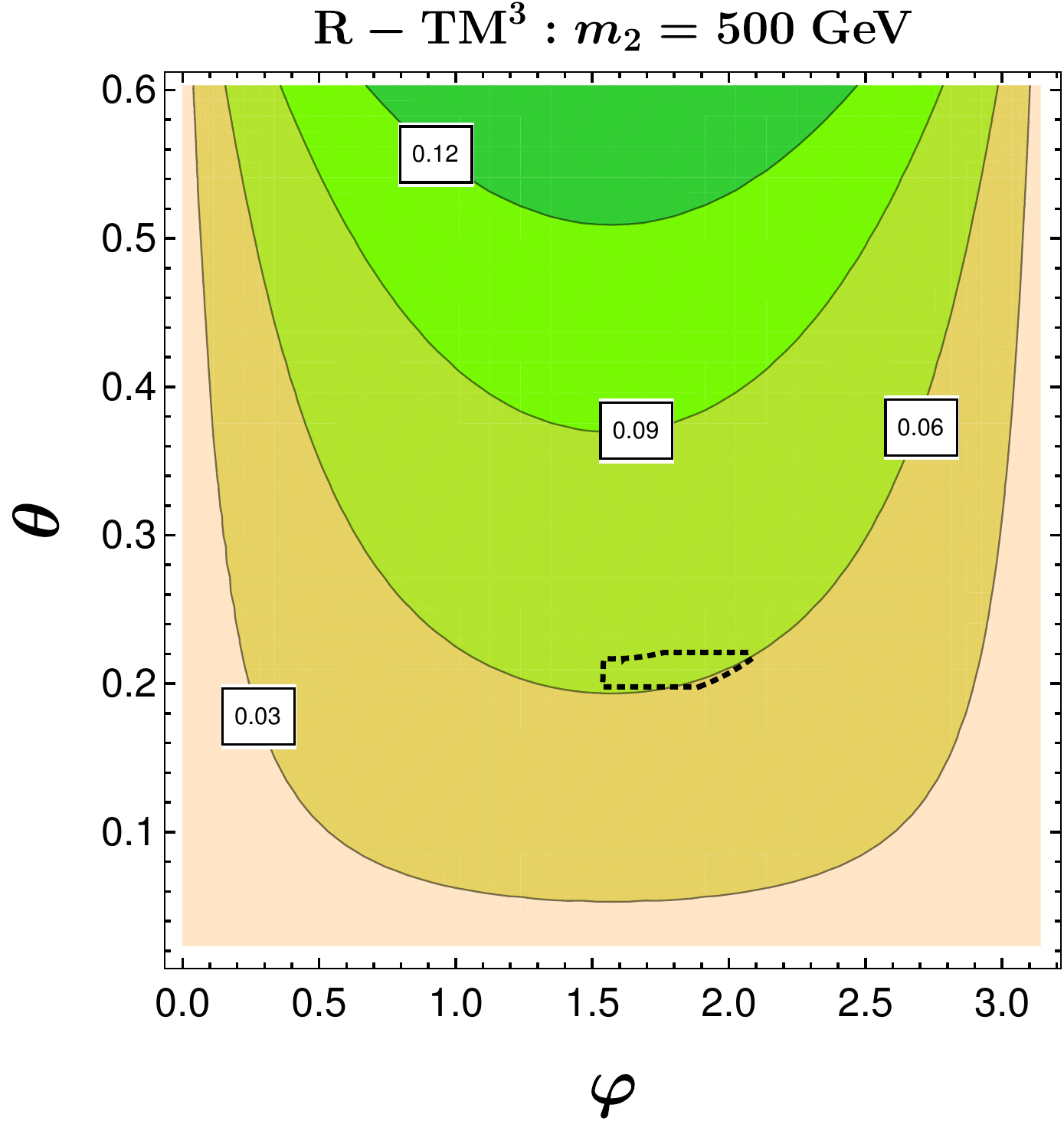}
	\caption{\label{R-TMLow2Up3_Contour} Compatibility between the freeze-in DN and the neutrino oscillation observables for patterns $\text{TM}_2$ and $\text{TM}^3$ with a retarded-cutting scheme. The area enclosed by the black-dotted line represents the $3\sigma$ allowed range of $\vert U\vert$ and a maximal CP measure $\mathcal{J}=\mathcal{J}_{CP}^{max}<0$. The contours denote the variations of $v_2$ (in unit of GeV), with each contour corresponding to the best-fit point of $Y_{\Delta B}$ given by eq.~\eqref{BAU today}.}
\end{figure}

It can be seen from eqs.~\eqref{Jarlskog} and \eqref{baryon asym in pattern-1}-\eqref{baryon asym in pattern-4} that, the baryon-number asymmetry $Y_{\Delta B}^{\text{R}}$ in the pattern $\text{TM}_{1}$ has an opposite sign to that in the pattern $\text{TM}_{2}$, while the CP measure $\mathcal{J}$ has the same sign in both patterns. This indicates that, to generate a positive $Y_{\Delta B}$, the product of the trigonometric functions, $\sin(2\theta)\sin\varphi$, should be negative (positive) for $\text{TM}_1$ ($\text{TM}_2$). However, if we follow the favored Dirac CP phase $\delta=[\pi,2\pi]$, which indicates a negative $\mathcal{J}$, the same factor $\sin(2\theta)\sin\varphi$ should be positive in both patterns. Therefore, for a successful DN, the pattern $\text{TM}_1$ is already disfavored by the neutrino oscillation data with a Dirac CP phase in the range $\delta=[\pi,2\pi]$.   For the patterns $\text{TM}^{2,3}$, on the other hand, $Y_{\Delta B}^{\text{R}}$ has basically the same value, while $\mathcal{J}$ has the opposite sign. This implies that the pattern $\text{TM}^2$ is also disfavored by the range of Dirac CP phase in realizing a successful DN. To visualize the sign significance observed above, we show in figure~\ref{R-TM-Angle} the allowed regions for the two mixing parameters, $(\theta,\varphi)$, under the individual constraint from a positive baryon-number asymmetry, a negative CP measure in neutrino oscillations, as well as the PMNS matrix element $\vert U_{13}\vert$. It should be emphasized here that the above conclusions and the yellow regions shown in figure~\ref{R-TM-Angle} are independent of the mass parameter $m_2$ chosen, though it was fixed at $500$~GeV for a simple estimation. For the two allowed patterns, $\text{R}\text{--}\text{TM}_2$ and $\text{R}\text{--}\text{TM}^3$, we further investigate in detail the compatibility between the freeze-in DN and the neutrino oscillation observables in a particular quadrant with $\theta=[0,\pi/2]$, which is shown in figure~\ref{R-TMLow2Up3_Contour}. 

\begin{figure}[ht]
	\centering
	\includegraphics[width=0.46\textwidth]{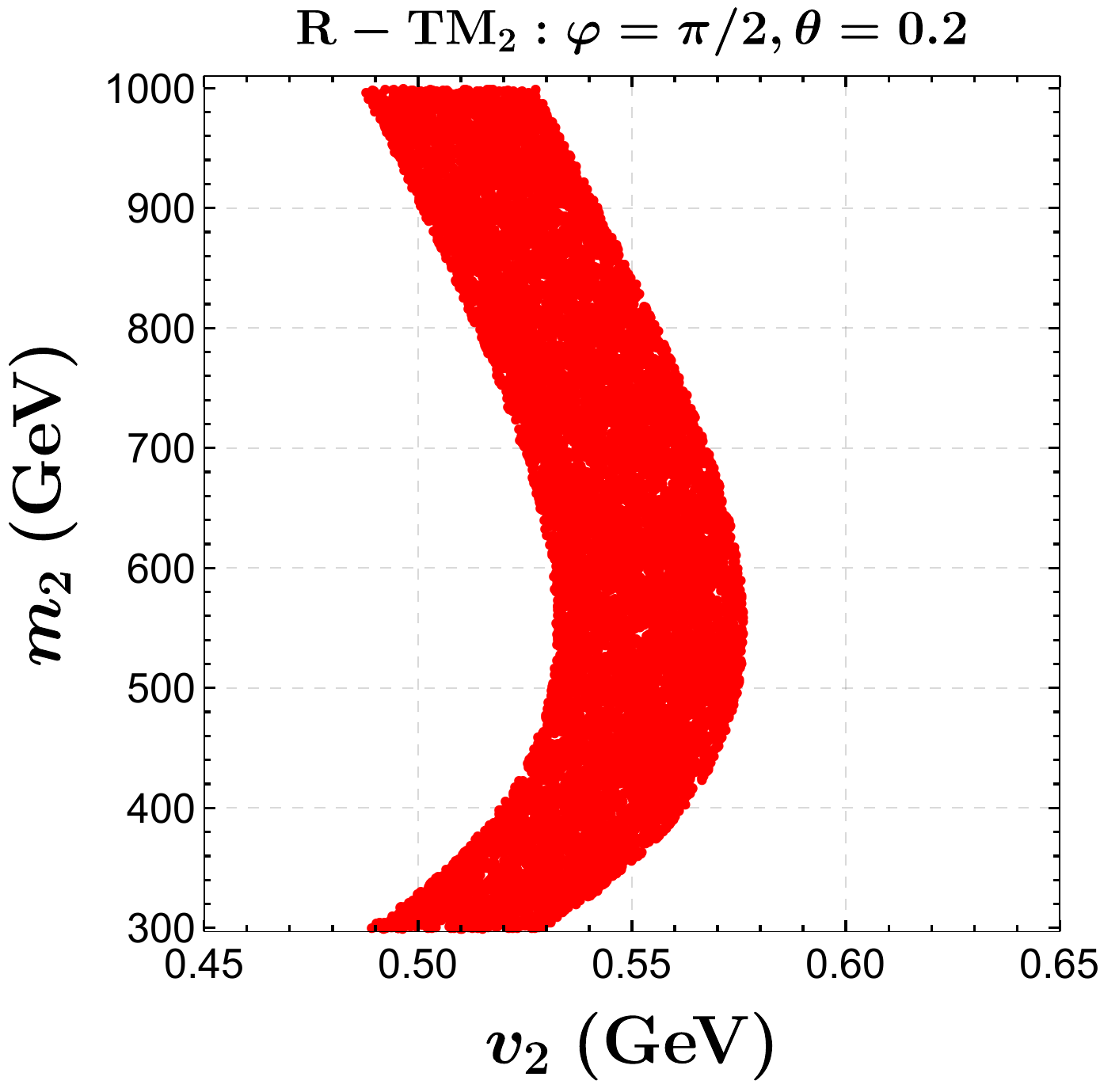}\qquad
	\includegraphics[width=0.46\textwidth]{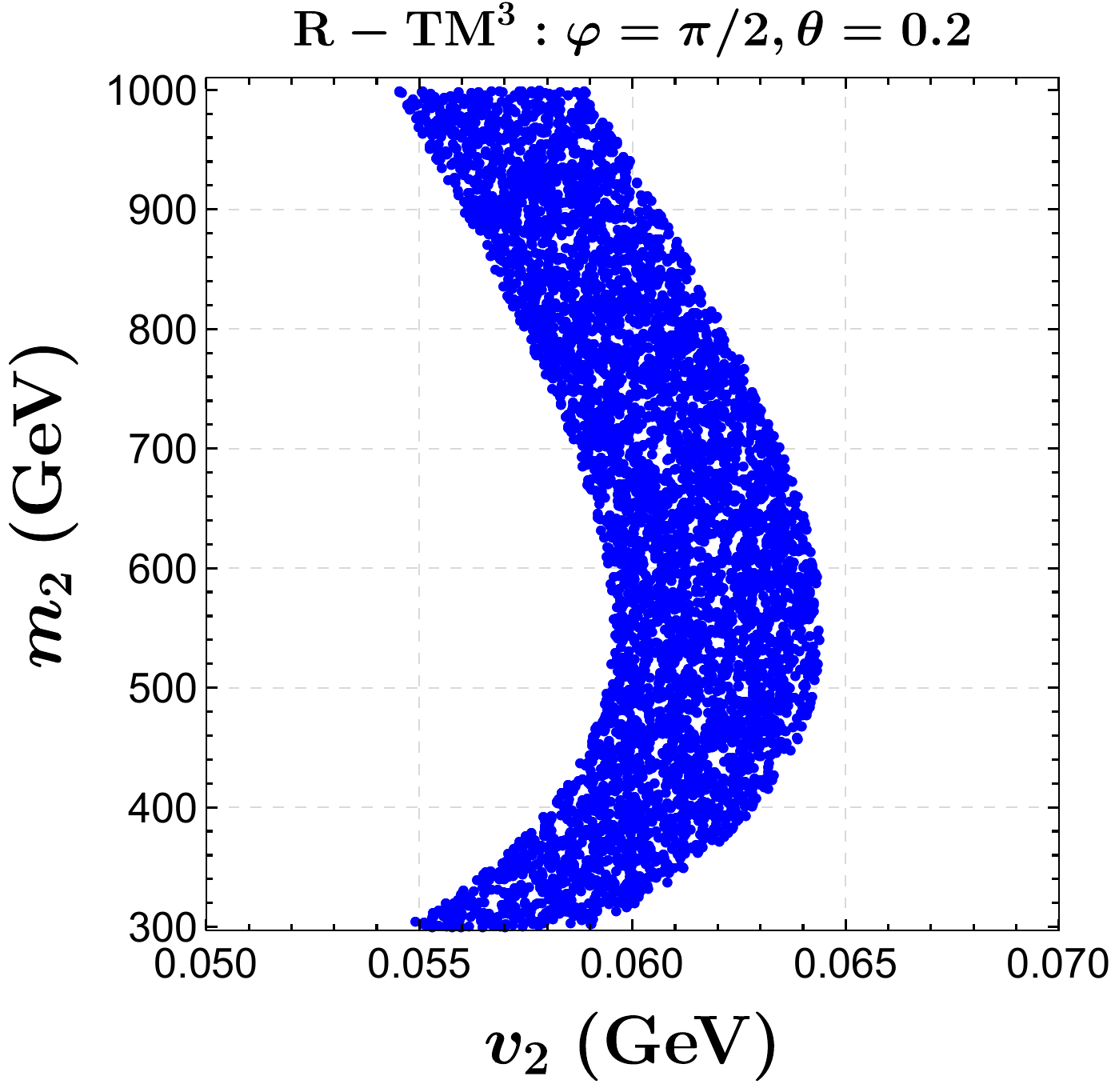}
	\caption{\label{v2m2-dependence} The $(v_2,m_2)$ dependence of the baryon-number asymmetry $Y^R_{\Delta B}$ with the preferred mixing parameters $\varphi=\pi/2$ and $\theta=0.2$ obtained from figure~\ref{R-TMLow2Up3_Contour}.}
\end{figure}

As can be seen from figure~\ref{R-TMLow2Up3_Contour}, both the patterns $\text{TM}_2$ and $\text{TM}^3$ can produce the $3\sigma$-allowed range of $\vert U\vert$ as well as a maximal CP measure $\mathcal{J}=\mathcal{J}_{CP}^{max}<0$ (the area enclosed by the black-dotted line), pointing out a Cabibbo-like angle $\theta\simeq 0.2$. It is also observed that the maximal leptonic CP asymmetry at low energy favors a maximal CP phase $\varphi\simeq \pi/2$, which is necessary for a lepton-number asymmetry at the high-temperature regime. On the other hand, since the mixing parameters $(\varphi, \theta)$ are determined by fitting the neutrino oscillation observables, a suggestive dependence of the baryon-number asymmetry on the mass parameter $m_2$ and the VEV $v_2$ can be derived by pinning down the preferred values of $(\varphi, \theta)$ obtained from figure~\ref{R-TMLow2Up3_Contour}. To this end, we vary the mass parameter $m_2$ in the interval $[300,1000]$~GeV, and fix the preferred Cabibbo-like angle $\theta\simeq 0.2$ and a maximal CP phase $\varphi\simeq \pi/2$ obtained from figure~\ref{R-TMLow2Up3_Contour}. Then, we show in figure~\ref{v2m2-dependence} the region of $(v_2, m_2)$ in which the 3$\sigma$-allowed range of $Y_{\Delta B}$ can be produced. It can be seen from the figure that the $m_2$ dependence is similar in both patterns, except that the required $v_2$ in $\text{TM}_2$, $0.49~\text{GeV}\lesssim v_2\lesssim 0.58~\text{GeV}$, is one order of magnitude larger than that in $\text{TM}^3$, $0.055~\text{GeV}\lesssim v_2 \lesssim0.064~\text{GeV}$.

For the preferred ranges of $v_2$, the Dirac neutrino Yukawa couplings are then estimated to be $Y_\nu\simeq \mathcal{O}(10^{-10})$ ($\text{TM}_2$) and $Y_\nu\simeq \mathcal{O}(10^{-9})$ ($\text{TM}^3$) for the neutrino masses at $\mathcal{O}(0.05)$~eV. This implies that the feeble neutrino Yukawa couplings of $\mathcal{O}(10^{-9}\text{--}10^{-10})$ obtained in the neutrinophilic 2HDM can account for the smallness of Dirac neutrino masses in a simple, but less aesthetic manner. We have also shown that it is just the feebleness that renders the accumulation of lepton-number asymmetry to convert into the baryon-number asymmetry via the rapid sphaleron transitions in the early Universe. 

Finally, we would like to discuss briefly possible experimental tests of the neutrinophilic 2HDM considered here to explain the BAU. In fact, the phenomenology and possible future tests of the scalar sector in the original neutrinophilic 2HDM have been extensively studied in ref.~\cite{Davidson:2009ha}. However, since much larger neutrino Yukawa couplings---seven to eight orders of magnitude bigger than needed here--were considered in the framework, not all the experimental proposals suggested in ref.~\cite{Davidson:2009ha} can be applied directly to our case. Due to the feeble neutrino Yukawa couplings, constraints from big bang nucleosynthesis can be readily satisfied, and indirect searches of the charged Higgs bosons through measurements of the lepton-flavor-violating decays $\ell \to \ell' \gamma$, the muon anomalous magnetic moment $(g-2)_{\mu}$, as well as the tree-level $\mu$ and $\tau$ decays $\ell \to \ell' \nu\bar{\nu}$ would be less possible. However, since we have invoked a new electroweak-scale Higgs doublet to implement the freeze-in DN, experimental tests of the additional scalars at current and future colliders are still feasible.
	
In this context, the charged Higgs $H^{\pm}$ can be pair-produced via $pp\to \gamma^*, Z^*\to H^+H^-$ at the LHC, and via $e^-e^+\to \gamma^*, Z^*\to H^+H^-$ at the future International Linear Collider (ILC)~\cite{Baer:2013cma} and Compact Linear Collider (CLIC)~\cite{Linssen:2012hp}. Because of the quasi-degenerate Higgs mass spectrum in our case, $H^{\pm}$ can decay mainly to $W^{\pm}$ and the SM Higgs $h$ at tree level, provided that $M_{H^\pm}>m_W+m_h$. Therefore, searching for the charged Higgs bosons through pair production at these colliders can be promising. The pseudoscalar $A$, on the other hand, can be pair-produced through loop-level photon-photon fusion at the ILC and CLIC. Also due to the quasi-degenerate Higgs mass spectrum, $A$ would decay predominantly to $Z$ and $h$, provided that $M_A> m_Z+m_h$. Thus, the collider search for $A$ is possible as well. As for the neutral Higgs $H$, it can be produced through vector boson fusion, \textit{e.g.},  $W^+W^-\to H$ or $ZZ\to H$. If $M_H>2m_W, 2m_Z$, it would decay mostly to $W^-W^+$ and $ZZ$. A more important remark on the two neutral Higgs bosons is that their decay channels are quite different from that in other 2HDM scenarios where $H, A$ can also decay to the SM fermion pairs~\cite{Branco:2011iw}. Hence, such a neutrinophilic 2HDM implemented in the freeze-in DN mechanism can be distinguished from the original version~\cite{Davidson:2009ha,Machado:2015sha,Bertuzzo:2015ada}, as well as from some popular 2HDM models (see, \textit{e.g.}, ref.~\cite{Branco:2011iw}).

\section{Conclusion}
\label{sec:con}

We have demonstrated in this paper that, when both the thermal effects at high temperature and the non-diagonal Yukawa textures of charged leptons and neutrinos are considered, it is feasible to account for the matter-antimatter asymmetry of the Universe within a minimal freeze-in DN setup. While the SM Higgs cannot generate the observed baryon-number asymmetry in such a minimal setup, the second Higgs doublet of the neutrinophilic 2HDM, when being in equilibrium with the thermal bath, can realize the freeze-in DN. To establish a direct connection between the high-temperature leptonic CP asymmetry and the low-energy neutrino oscillation observables, we have considered various minimal corrections to the TB mixing pattern. It is found that the patterns with a Cabibbo-like mixing angle and a maximal CP-violating phase can produce compatible neutrino oscillation observables with a (negative) maximal CP measure and, at the same time, account for the matter-antimatter asymmetry of the Universe observed today.

Such a minimal setup realized in this paper is predictable on account of the correlation between the BAU and the neutrino oscillation observables, and might also be testable at current and future colliders in terms of the electroweak scalars introduced to generate the neutrino masses and to implement the BAU.

\section*{Acknowledgements}
This work is supported by the National Natural Science Foundation of China under Grant Nos.~12075097, 11675061 and 11775092, as well as by the Fundamental Research Funds for the Central Universities under Grant Nos.~CCNU20TS007 and 2019YBZZ079.

\bibliographystyle{JHEP}
\bibliography{reference}

\providecommand{\href}[2]{#2}\begingroup\raggedright\begin{thebibliography}{10}

\bibitem{Hall:2009bx}
L.~J. Hall, K.~Jedamzik, J.~March-Russell, and S.~M. West, {\it {Freeze-In
  Production of FIMP Dark Matter}},  {\it JHEP} {\bf 03} (2010) 080,
  [\href{http://arxiv.org/abs/0911.1120}{{\tt arXiv:0911.1120}}].

\bibitem{Bernal:2017kxu}
N.~Bernal, M.~Heikinheimo, T.~Tenkanen, K.~Tuominen, and V.~Vaskonen, {\it {The
  Dawn of FIMP Dark Matter: A Review of Models and Constraints}},  {\it Int. J.
  Mod. Phys.} {\bf A32} (2017), no.~27 1730023,
  [\href{http://arxiv.org/abs/1706.07442}{{\tt arXiv:1706.07442}}].

\bibitem{Dick:1999je}
K.~Dick, M.~Lindner, M.~Ratz, and D.~Wright, {\it {Leptogenesis with Dirac
  neutrinos}},  {\it Phys. Rev. Lett.} {\bf 84} (2000) 4039--4042,
  [\href{http://arxiv.org/abs/hep-ph/9907562}{{\tt hep-ph/9907562}}].

\bibitem{Kuzmin:1985mm}
V.~A. Kuzmin, V.~A. Rubakov, and M.~E. Shaposhnikov, {\it {On the Anomalous
  Electroweak Baryon Number Nonconservation in the Early Universe}},  {\it
  Phys. Lett.} {\bf 155B} (1985) 36.

\bibitem{Murayama:2002je}
H.~Murayama and A.~Pierce, {\it {Realistic Dirac leptogenesis}},  {\it Phys.
  Rev. Lett.} {\bf 89} (2002) 271601,
  [\href{http://arxiv.org/abs/hep-ph/0206177}{{\tt hep-ph/0206177}}].

\bibitem{Cerdeno:2006ha}
D.~G. Cerdeno, A.~Dedes, and T.~E.~J. Underwood, {\it {The Minimal Phantom
  Sector of the Standard Model: Higgs Phenomenology and Dirac Leptogenesis}},
  {\it JHEP} {\bf 09} (2006) 067,
  [\href{http://arxiv.org/abs/hep-ph/0607157}{{\tt hep-ph/0607157}}].

\bibitem{Gu:2007mc}
P.-H. Gu, H.-J. He, and U.~Sarkar, {\it {Realistic neutrinogenesis with
  radiative vertex correction}},  {\it Phys. Lett.} {\bf B659} (2008) 634--639,
  [\href{http://arxiv.org/abs/0709.1019}{{\tt arXiv:0709.1019}}].

\bibitem{Bechinger:2009qk}
A.~Bechinger and G.~Seidl, {\it {Resonant Dirac leptogenesis on throats}},
  {\it Phys. Rev.} {\bf D81} (2010) 065015,
  [\href{http://arxiv.org/abs/0907.4341}{{\tt arXiv:0907.4341}}].

\bibitem{Narendra:2017uxl}
N.~Narendra, N.~Sahoo, and N.~Sahu, {\it {Dark matter assisted Dirac
  leptogenesis and neutrino mass}},  {\it Nucl. Phys.} {\bf B936} (2018)
  76--90, [\href{http://arxiv.org/abs/1712.02960}{{\tt arXiv:1712.02960}}].

\bibitem{Giudice:2003jh}
G.~F. Giudice, A.~Notari, M.~Raidal, A.~Riotto, and A.~Strumia, {\it {Towards a
  complete theory of thermal leptogenesis in the SM and MSSM}},  {\it Nucl.
  Phys.} {\bf B685} (2004) 89--149,
  [\href{http://arxiv.org/abs/hep-ph/0310123}{{\tt hep-ph/0310123}}].

\bibitem{Buchmuller:2004nz}
W.~Buchmuller, P.~Di~Bari, and M.~Plumacher, {\it {Leptogenesis for
  pedestrians}},  {\it Annals Phys.} {\bf 315} (2005) 305--351,
  [\href{http://arxiv.org/abs/hep-ph/0401240}{{\tt hep-ph/0401240}}].

\bibitem{Davidson:2008bu}
S.~Davidson, E.~Nardi, and Y.~Nir, {\it {Leptogenesis}},  {\it Phys. Rept.}
  {\bf 466} (2008) 105--177, [\href{http://arxiv.org/abs/0802.2962}{{\tt
  arXiv:0802.2962}}].

\bibitem{Harrison:2002er}
P.~F. Harrison, D.~H. Perkins, and W.~G. Scott, {\it {Tri-bimaximal mixing and
  the neutrino oscillation data}},  {\it Phys. Lett.} {\bf B530} (2002) 167,
  [\href{http://arxiv.org/abs/hep-ph/0202074}{{\tt hep-ph/0202074}}].

\bibitem{Albright:2008rp}
C.~H. Albright and W.~Rodejohann, {\it {Comparing Trimaximal Mixing and Its
  Variants with Deviations from Tri-bimaximal Mixing}},  {\it Eur. Phys. J.}
  {\bf C62} (2009) 599--608, [\href{http://arxiv.org/abs/0812.0436}{{\tt
  arXiv:0812.0436}}].

\bibitem{He:2011gb}
X.-G. He and A.~Zee, {\it {Minimal Modification to Tri-bimaximal Mixing}},
  {\it Phys. Rev.} {\bf D84} (2011) 053004,
  [\href{http://arxiv.org/abs/1106.4359}{{\tt arXiv:1106.4359}}].

\bibitem{Hambye:2016sby}
T.~Hambye and D.~Teresi, {\it {Higgs doublet decay as the origin of the baryon
  asymmetry}},  {\it Phys. Rev. Lett.} {\bf 117} (2016), no.~9 091801,
  [\href{http://arxiv.org/abs/1606.00017}{{\tt arXiv:1606.00017}}].

\bibitem{Das1997}
A.~Das, {\it Finite Temperature Field Theory}.
\newblock WORLD SCIENTIFIC, 1997.

\bibitem{Harvey:1990qw}
J.~A. Harvey and M.~S. Turner, {\it {Cosmological baryon and lepton number in
  the presence of electroweak fermion number violation}},  {\it Phys. Rev. D}
  {\bf 42} (1990) 3344--3349.

\bibitem{Landsman:1986uw}
N.~P. Landsman and C.~G. van Weert, {\it {Real and Imaginary Time Field Theory
  at Finite Temperature and Density}},  {\it Phys. Rept.} {\bf 145} (1987) 141.

\bibitem{Weldon:1982bn}
H.~A. Weldon, {\it {Effective Fermion Masses of Order gT in High Temperature
  Gauge Theories with Exact Chiral Invariance}},  {\it Phys. Rev.} {\bf D26}
  (1982) 2789.

\bibitem{Garny:2009rv}
M.~Garny, A.~Hohenegger, A.~Kartavtsev, and M.~Lindner, {\it {Systematic
  approach to leptogenesis in nonequilibrium QFT: Vertex contribution to the
  CP-violating parameter}},  {\it Phys. Rev. D} {\bf 80} (2009) 125027,
  [\href{http://arxiv.org/abs/0909.1559}{{\tt arXiv:0909.1559}}].

\bibitem{Garny:2009qn}
M.~Garny, A.~Hohenegger, A.~Kartavtsev, and M.~Lindner, {\it {Systematic
  approach to leptogenesis in nonequilibrium QFT: Self-energy contribution to
  the CP-violating parameter}},  {\it Phys. Rev. D} {\bf 81} (2010) 085027,
  [\href{http://arxiv.org/abs/0911.4122}{{\tt arXiv:0911.4122}}].

\bibitem{Garny:2010nj}
M.~Garny, A.~Hohenegger, and A.~Kartavtsev, {\it {Medium corrections to the
  CP-violating parameter in leptogenesis}},  {\it Phys. Rev.} {\bf D81} (2010)
  085028, [\href{http://arxiv.org/abs/1002.0331}{{\tt arXiv:1002.0331}}].

\bibitem{Kobes:1986za}
R.~L. Kobes and G.~W. Semenoff, {\it {Discontinuities of Green Functions in
  Field Theory at Finite Temperature and Density. 2}},  {\it Nucl. Phys.} {\bf
  B272} (1986) 329--364.

\bibitem{Kobes:1990ua}
R.~Kobes, {\it {Retarded functions, dispersion relations, and Cutkosky rules at
  zero and finite temperature}},  {\it Phys. Rev.} {\bf D43} (1991) 1269--1282.

\bibitem{Gelis:1997zv}
F.~Gelis, {\it {Cutting rules in the real time formalisms at finite
  temperature}},  {\it Nucl. Phys.} {\bf B508} (1997) 483--505,
  [\href{http://arxiv.org/abs/hep-ph/9701410}{{\tt hep-ph/9701410}}].

\bibitem{Kobes:1990kr}
R.~Kobes, {\it {A Correspondence Between Imaginary Time and Real Time Finite
  Temperature Field Theory}},  {\it Phys. Rev.} {\bf D42} (1990) 562--572.

\bibitem{Frossard:2012pc}
T.~Frossard, M.~Garny, A.~Hohenegger, A.~Kartavtsev, and D.~Mitrouskas, {\it
  {Systematic approach to thermal leptogenesis}},  {\it Phys. Rev.} {\bf D87}
  (2013), no.~8 085009, [\href{http://arxiv.org/abs/1211.2140}{{\tt
  arXiv:1211.2140}}].

\bibitem{Branco:2002kt}
G.~C. Branco, R.~Gonzalez~Felipe, F.~R. Joaquim, and M.~N. Rebelo, {\it
  {Leptogenesis, CP violation and neutrino data: What can we learn?}},  {\it
  Nucl. Phys.} {\bf B640} (2002) 202--232,
  [\href{http://arxiv.org/abs/hep-ph/0202030}{{\tt hep-ph/0202030}}].

\bibitem{Altarelli:2010gt}
G.~Altarelli and F.~Feruglio, {\it {Discrete Flavor Symmetries and Models of
  Neutrino Mixing}},  {\it Rev. Mod. Phys.} {\bf 82} (2010) 2701--2729,
  [\href{http://arxiv.org/abs/1002.0211}{{\tt arXiv:1002.0211}}].

\bibitem{Xing:2019vks}
Z.-z. Xing, {\it {Flavor structures of charged fermions and massive
  neutrinos}},  {\it Phys. Rept.} {\bf 854} (2020) 1--147,
  [\href{http://arxiv.org/abs/1909.09610}{{\tt arXiv:1909.09610}}].

\bibitem{Feruglio:2019ktm}
F.~Feruglio and A.~Romanino, {\it {Neutrino Flavour Symmetries}},
  \href{http://arxiv.org/abs/1912.06028}{{\tt arXiv:1912.06028}}.

\bibitem{Esteban:2016qun}
I.~Esteban, M.~C. Gonzalez-Garcia, M.~Maltoni, I.~Martinez-Soler, and
  T.~Schwetz, {\it {Updated fit to three neutrino mixing: exploring the
  accelerator-reactor complementarity}},  {\it JHEP} {\bf 01} (2017) 087,
  [\href{http://arxiv.org/abs/1611.01514}{{\tt arXiv:1611.01514}}].

\bibitem{deSalas:2017kay}
P.~F. de~Salas, D.~V. Forero, C.~A. Ternes, M.~Tortola, and J.~W.~F. Valle,
  {\it {Status of neutrino oscillations 2018: 3$\sigma$ hint for normal mass
  ordering and improved CP sensitivity}},  {\it Phys. Lett.} {\bf B782} (2018)
  633--640, [\href{http://arxiv.org/abs/1708.01186}{{\tt arXiv:1708.01186}}].

\bibitem{Esteban:2018azc}
I.~Esteban, M.~C. Gonzalez-Garcia, A.~Hernandez-Cabezudo, M.~Maltoni, and
  T.~Schwetz, {\it {Global analysis of three-flavour neutrino oscillations:
  synergies and tensions in the determination of $\theta_{23}$, $\delta_{CP}$,
  and the mass ordering}},  {\it JHEP} {\bf 01} (2019) 106,
  [\href{http://arxiv.org/abs/1811.05487}{{\tt arXiv:1811.05487}}].

\bibitem{Capozzi:2018ubv}
F.~Capozzi, E.~Lisi, A.~Marrone, and A.~Palazzo, {\it {Current unknowns in the
  three neutrino framework}},  {\it Prog. Part. Nucl. Phys.} {\bf 102} (2018)
  48--72, [\href{http://arxiv.org/abs/1804.09678}{{\tt arXiv:1804.09678}}].

\bibitem{deSalas:2020pgw}
P.~de~Salas, D.~Forero, S.~Gariazzo, P.~Martínez-Miravé, O.~Mena, C.~Ternes,
  M.~Tórtola, and J.~Valle, {\it {2020 Global reassessment of the neutrino
  oscillation picture}},  \href{http://arxiv.org/abs/2006.11237}{{\tt
  arXiv:2006.11237}}.

\bibitem{Esteban:2020cvm}
I.~Esteban, M.~Gonzalez-Garcia, M.~Maltoni, T.~Schwetz, and A.~Zhou, {\it {The
  fate of hints: updated global analysis of three-flavor neutrino
  oscillations}},  {\it JHEP} {\bf 09} (2020) 178,
  [\href{http://arxiv.org/abs/2007.14792}{{\tt arXiv:2007.14792}}].

\bibitem{Jarlskog:1985ht}
C.~Jarlskog, {\it {Commutator of the Quark Mass Matrices in the Standard
  Electroweak Model and a Measure of Maximal CP Violation}},  {\it Phys.\ Rev.\
  Lett.} {\bf 55} (1985) 1039.

\bibitem{Tanabashi:2018oca}
{\bf Particle Data Group} Collaboration, M.~Tanabashi et~al., {\it {Review of
  Particle Physics}},  {\it Phys. Rev.} {\bf D98} (2018), no.~3 030001.

\bibitem{Aghanim:2018eyx}
{\bf Planck} Collaboration, N.~Aghanim et~al., {\it {Planck 2018 results. VI.
  Cosmological parameters}},  {\it Astron. Astrophys.} {\bf 641} (2020) A6,
  [\href{http://arxiv.org/abs/1807.06209}{{\tt arXiv:1807.06209}}].

\bibitem{Davidson:2009ha}
S.~M. Davidson and H.~E. Logan, {\it {Dirac neutrinos from a second Higgs
  doublet}},  {\it Phys. Rev.} {\bf D80} (2009) 095008,
  [\href{http://arxiv.org/abs/0906.3335}{{\tt arXiv:0906.3335}}].

\bibitem{Machado:2015sha}
P.~A.~N. Machado, Y.~F. Perez, O.~Sumensari, Z.~Tabrizi, and R.~Z. Funchal,
  {\it {On the Viability of Minimal Neutrinophilic Two-Higgs-Doublet Models}},
  {\it JHEP} {\bf 12} (2015) 160, [\href{http://arxiv.org/abs/1507.07550}{{\tt
  arXiv:1507.07550}}].

\bibitem{Bertuzzo:2015ada}
E.~Bertuzzo, Y.~F. Perez~G., O.~Sumensari, and R.~Zukanovich~Funchal, {\it
  {Limits on Neutrinophilic Two-Higgs-Doublet Models from Flavor Physics}},
  {\it JHEP} {\bf 01} (2016) 018, [\href{http://arxiv.org/abs/1510.04284}{{\tt
  arXiv:1510.04284}}].

\bibitem{Cline:1995dg}
J.~M. Cline, K.~Kainulainen, and A.~P. Vischer, {\it {Dynamics of two Higgs
  doublet CP violation and baryogenesis at the electroweak phase transition}},
  {\it Phys.\ Rev.\ D} {\bf 54} (1996) 2451--2472,
  [\href{http://arxiv.org/abs/hep-ph/9506284}{{\tt hep-ph/9506284}}].

\bibitem{Li:2018aov}
S.-P. Li, X.-Q. Li, and Y.-D. Yang, {\it {Muon $g-2$ in a $U(1)$-symmetric
  Two-Higgs-Doublet Model}},  {\it Phys. Rev.} {\bf D99} (2019), no.~3 035010,
  [\href{http://arxiv.org/abs/1808.02424}{{\tt arXiv:1808.02424}}].

\bibitem{Baer:2013cma}
H.~Baer et~al., {\it {The International Linear Collider Technical Design Report
  - Volume 2: Physics}},  \href{http://arxiv.org/abs/1306.6352}{{\tt
  arXiv:1306.6352}}.

\bibitem{Linssen:2012hp}
L.~Linssen, A.~Miyamoto, M.~Stanitzki, and H.~Weerts, {\it {Physics and
  Detectors at CLIC: CLIC Conceptual Design Report}},
  \href{http://arxiv.org/abs/1202.5940}{{\tt arXiv:1202.5940}}.

\bibitem{Branco:2011iw}
G.~Branco, P.~Ferreira, L.~Lavoura, M.~Rebelo, M.~Sher, and J.~P. Silva, {\it
  {Theory and phenomenology of two-Higgs-doublet models}},  {\it Phys. Rept.}
  {\bf 516} (2012) 1--102, [\href{http://arxiv.org/abs/1106.0034}{{\tt
  arXiv:1106.0034}}].

\end{thebibliography}\endgroup

\end{document}